\documentclass[twocolumn,showpacs,preprintnumbers,amsmath,amssymb]{revtex4}

\usepackage{graphicx}
\usepackage{dcolumn}
\usepackage{bm}

\begin{document}


\title{Comprehensive proof of the Greenberger-Horne-Zeilinger
Theorem for the four-qubit system}
\author{Li Tang}
\email{tangli@wipm.ac.cn}
\affiliation{%
State Key Laboratory of Magnetic Resonance and Atomic and
Molecular Physics, Wuhan Institute of Physics and Mathematics,
Chinese Academy of Sciences, Wuhan 430071, China}%
\affiliation{%
Center for Cold Atom Physics, Chinese Academy of Sciences,
Wuhan 430071, China}%
\affiliation{%
Graduate School, Chinese Academy of Sciences, Wuhan
430071, China}%

\author{Jie Zhong}
\affiliation{%
Laboratory of Mathematical Physics, Wuhan Institute of Physics and
Mathematics, Chinese Academy of Sciences, Wuhan 430071, China}%
\affiliation{%
Graduate School, Chinese Academy of Sciences, Wuhan
430071, China}%

\author{Yaofeng Ren}
\affiliation{%
Department of Mathematics, The Naval University of Engineering,
Wuhan 430033, China}%

\author{Mingsheng Zhan}
\affiliation{%
State Key Laboratory of Magnetic Resonance and Atomic and
Molecular Physics, Wuhan Institute of Physics and Mathematics,
Chinese Academy of Sciences, Wuhan 430071, China}%
\affiliation{%
Center for Cold Atom Physics, Chinese Academy of Sciences,
Wuhan 430071, China}%

\author{Zeqian Chen}
\email{zqchen@wipm.ac.cn}
\affiliation{%
State Key Laboratory of Magnetic Resonance and Atomic and
Molecular Physics and Laboratory of Mathematical Physics, Wuhan
Institute of Physics and Mathematics, Chinese Academy of Sciences,
Wuhan 430071, China}%

\date{\today}

\begin{abstract}
Greenberger-Horne-Zeilinger (GHZ) theorem asserts that there is a
set of mutually commuting nonlocal observables with a common
eigenstate on which those observables assume values that refute
the attempt to assign values only required to have them by the
local realism of Einstein, Podolsky, and Rosen (EPR). It is known
that for a three-qubit system there is only one form of the
GHZ-Mermin-like argument with equivalence up to a local unitary
transformation, which is exactly Mermin's version of the GHZ
theorem. In this paper, however, for a four-qubit system which was
originally studied by GHZ, we show that there are nine distinct
forms of the GHZ-Mermin-like argument. The proof is obtained by
using some geometric invariants to characterize the sets of
mutually commuting nonlocal spin observables on the four-qubit
system. It is proved that there are at most nine elements (except
for a different sign) in a set of mutually commuting nonlocal spin
observables in the four-qubit system, and each GHZ-Mermin-like
argument involves a set of at least five mutually commuting
four-qubit nonlocal spin observables with a GHZ state as a common
eigenstate in GHZ's theorem. Therefore, we present a complete
construction of the GHZ theorem for the four-qubit system.
\end{abstract}

\maketitle

\section{Introduction }

Bell's inequality \cite{Bell} indicates that certain statistical
correlations predicted by quantum mechanics for measurements on
two-qubit ensembles cannot be understood within a realistic
picture based on Einstein, Podolsky, and Rosen's (EPR's) notion of
local realism \cite{EPR}. There is an unsatisfactory feature in
the derivation of Bell's inequality that such a local realistic
and, consequently, classical picture can explain perfect
correlations and is only in conflict with statistical prediction
of quantum mechanics. Strikingly enough, the
Greenberger-Horne-Zeilinger (GHZ's) theorem exhibits that the
contradiction between quantum mechanics and local realistic
theories arises even for definite predictions on a four-qubit
system \cite{GHZ}. Mermin \cite{M90} subsequently refined the
original GHZ argument on a three-qubit system. Let us recall that
their approaches were characterized by the following premises:

(a)  a set of mutually commuting nonlocal observables,

(b)  a common eigenstate on which those observables assume values
that refute the attempt to assign values only required to have
them by EPR's local realism.

Based on this criterion of a GHZ-Mermin-like argument, we define a
GHZ-Mermin experiment by a set of mutually commuting nonlocal
observables with at least two different observables at each site.
(Note that, a common local observable does not provide a random
selection of measurements and so plays no role in the
GHZ-Mermin-type proof.) A GHZ-Mermin experiment presenting a
GHZ-Mermin-like argument on a certain common eigenstate is said to
be nontrivial. There is no nontrivial GHZ-Mermin experiment in the
two-qubit system, while in the three-qubit system there is only
one nontrivial GHZ-Mermin experiment (with equivalence up to a
local unitary transformation), which is exactly Mermin's version
of the GHZ theorem \cite{Chen}. On the other hand, the
GHZ-Mermin-like argument has been extended to $n$ qubits
\cite{PRC-C}, and to multiparty multilevel systems \cite{Cabello}.
So far, however, no complete construction of nontrivial GHZ-Mermin
experiments is presented beyond the three-qubit system as noted in
\cite{Chen}, there are only partial results \cite{R-Z}.

In this paper, we will construct all nontrivial GHZ-Mermin
experiments of the four-qubit system, for which the GHZ-like
argument was developed originally by GHZ \cite{GHZ}. It is proved
that there are nine distinct forms of the GHZ-Mermin-like argument
on the four-qubit system, and each GHZ-Mermin-like argument
involves a set of at least five mutually commuting four-qubit
nonlocal spin observables with a GHZ state as a common eigenstate
in GHZ's theorem. Precisely, we obtain the following results.

(i)~~All four-qubit GHZ-Mermin experiments of at most four
elements are trivial.

(ii)~~Four-qubit GHZ-Mermin experiments of five (6, 7, or 8)
elements possess 11 (9, 5, or 3) different forms, two of which are
nontrivial.

(iii)~~A four-qubit GHZ-Mermin experiment contains at most nine
elements and, the experiments of nine elements have two different
forms, one of which is trivial, while another one is nontrivial.

(iv)~~In every nontrivial GHZ-Mermin experiment for the four-qubit
system, the associated states exhibiting an ``all versus nothing"
contradiction between quantum mechanics and $\mathrm{EPR}$'s local
realism must be $\mathrm{GHZ}$ states.

Our proof is based on some subtle mathematical arguments. We first
classify the equivalence of GHZ-Mermin experiments by two basic
symmetries acting on them. Then, we define two geometric
invariants for a GHZ-Mermin experiment, which can be used to
distinguish two inequivalent experiments. These arguments can be
easily extended to $n$ qubits.

The structure of this paper is as follows. In Sec.II, we first
prove a lemma on the structure of two commuting nonlocal spin
observables of $n$ qubits. Then, we discuss two basic symmetries
($\mathrm{(S_1)}$ and $\mathrm{(S_2)}$) acting on GHZ-Mermin
experiments of $n$ qubits. By these two basic symmetries we define
the equivalence of GHZ-Mermin experiments. We illustrate that a
four-qubit GHZ-Mermin experiment of three elements must
equivalently be one of three different forms. Finally, we define
two geometric invariants ($\mathrm{C}$-invariants and
$\mathrm{R}$-invariants) for a GHZ-Mermin experiment. These two
geometric invariants are invariant under $\mathrm{(S_1)},
\mathrm{(S_2)},$ and local unitary transformations
($\mathrm{LU}$). They paly a crucial role in the equivalence of
GHZ-Mermin experiments. In Sec.III, we show that a four-qubit
GHZ-Mermin experiment of four elements must equivalently be one of
seven different forms. Then we prove that every four-qubit
GHZ-Mermin experiment of three or four elements is trivial. In
Sec.IV, we show that four-qubit GHZ-Mermin experiments of five (6,
7, or 8) elements possess 11 (9, 5, or 3) different forms. It is
proved that in each case there are two nontrivial GHZ-Mermin
experiments and, the associated states exhibiting an ``all versus
nothing" contradiction between quantum mechanics and
$\mathrm{EPR}$'s local realism are $\mathrm{GHZ}$ states. In
Sec.V, we prove that a four-qubit GHZ-Mermin experiment contains
at most nine elements and the experiments of nine elements have
two different forms, one of which is trivial while another one is
nontrivial. Finally, in Sec.VI we give some concluding remarks and
questions for further consideration.

\section{GHZ-Mermin Experiments and symmetries}

Let us consider a system of $n$ qubits labelled  by
$1,2,\cdots,n.$ Let $A_j, A'_j$ denote spin observables on the
$j$th qubit, $j = 1, 2, \cdots, n.$ For $A^{(\prime)}_j=
\vec{a}^{(\prime)}_j \cdot \vec{\sigma}_j$ $(1\leq j \leq n),$ we
write$$(A_j,A'_j) = (\vec{a}_j, \vec{a}'_j ), A_j \times A'_j = (
\vec{a}_j \times \vec{a}'_j) \cdot \vec{\sigma}_j.$$Here
$\vec{\sigma}_j = (\sigma^j_x, \sigma^j_y, \sigma^j_z )$ are the
Pauli matrices for the $j$th qubit; the vectors
$\vec{a}^{(\prime)}_j$ are all unit vectors in $\mathbb{R}^3.$ It
is easy to check that\begin{equation}A_j A'_j = (A_j,A'_j) + i A_j
\times A'_j, \tag{2.1}\label{eq:2-1}\end{equation}\begin{equation}
A'_jA_j = (A_j, A'_j) - i A_j \times A'_j,
\tag{2.2}\label{eq:2-2}\end{equation}\begin{equation}\| A_j \times
A'_j \|^2=1- (A_j, A'_j)^2. \tag{2.3}\label{eq:2-3}
\end{equation}Also, $A_j \times A'_j = 0$ if and only if
$A_j = \pm A'_j,$ i.e., $A_j$ is parallel to $A'_j;$ $(A_j,A'_j) =
0$ if and only if $A_j$ is orthogonal to $A'_j,$ denoted by $A_j
\perp A'_j.$

We write $A^{(\prime)}_1 \cdots A^{(\prime)}_n,$ etc., as
shorthand for $A^{(\prime)}_1 \otimes \cdots \otimes
A^{(\prime)}_n.$ The following lemma clarifies the inner structure
of mutually commuting nonlocal spin observables of the $n$-qubit
system.

{\it Lemma: Two nonlocal $n$-qubit spin observables $A_1 \cdots
A_n$ and $A'_1 \cdots A'_n$ are commuting if and only if for every
$j = 1, 2, \cdots, n,$ $A_j$ is either  parallel or orthogonal to
$A'_j,$ and the number of sites at which the corresponding local
spin observables are orthogonal to each other is even.}

{\it Proof.}~~The sufficiency is clear. Indeed, by Eqs.(2.1) and
(2.2), we have that $A_j A'_j = - A'_j A_j$ whenever $(A_j,A'_j) =
0.$ Since the number of elements of $\{j: A_j \perp A'_j \}$ is
even, it is immediately concluded that $A_1 \cdots A_n$ and $A'_1
\cdots A'_n$ are commuting.

To prove the necessity, suppose that $A_1 \cdots A_n$ and $A'_1
\cdots A'_n$ are commuting. For every unit vector $|u_1 \rangle
\otimes \cdots \otimes |u_n \rangle,$ one has\begin{equation*}
\prod_{j = 1}^n \| A_j A'_j |u_j \rangle \|^2 = \prod_{j = 1}^n
\langle A_j A'_j u_j |A'_j A_j u_j \rangle.\end{equation*}By
Eqs.(2.1) and (2.2), we have
\begin{eqnarray*}
\|A_j A'_j |u_j \rangle\|^{2} = (A_j, A'_j )^2 + \|A_j \times A'_j
|u_j \rangle\|^2,\end{eqnarray*}
\begin{eqnarray*} \langle A_j A'_j
u_j |A'_j A_j u_j \rangle = (A_j, A'_j )^2-\|A_j \times A'_j |u_j
\rangle \|^2
\\[0.4cm]
- 2i (A_j, A'_j ) \langle u_j |A_j \times A'_j | u_j \rangle.
\end{eqnarray*}Note that, if $A_j \times A'_j \neq 0,$ there correspond to two
eigenvalues $\pm \|A_j \times A'_j \|$ with the corresponding unit
eigenvectors $| u^{\pm}_j \rangle.$ In this case, we set $ | u_j
\rangle = ( | u^+_j \rangle + | u^-_j \rangle ) / \sqrt{2}$ and
obtain $\langle u_j |A_j \times A'_j |u_j \rangle = 0.$ Hence, we
have\begin{eqnarray*}\label{l} \prod_{j=1}^n \left [ (A_j, A'_j
)^{2} + \|A_j \times A'_j |u_j \rangle \|^2 \right ]\\[0.4cm]
= \prod_{j = 1}^n \left [ (A_j, A'_j )^2 -\|A_j \times A'_j |u_j
\rangle \|^2 \right ].\end{eqnarray*}
This immediately concludes
that either $A_j \times A'_j = 0$ or $(A_j,A'_j) = 0$ for each $j
= 1, 2, \cdots, n.$ On the other hand, by Eqs.(2.1) and (2.2) we
have that $A_j A'_j = - A'_j A_j$ whenever $(A_j,A'_j) = 0.$
Therefore, the number of elements of $A_j \perp A'_j$ is even.

The Lemma tells us that two commuting nonlocal spin observables of
the $n$-qubit system have a nice structure, which has been used to
clarify the geometric structure of GHZ-Mermin experiments of both
two-qubit and three-qubit systems in \cite{Chen}. For convenience,
we reformulate the Lemma in the case of four qubits that two
four-qubit nonlocal spin observables $A_1 A_2 A_3 A_4$ and $A'_1
A'_2 A'_3 A'_4$ are commuting if and only if one of the following
conditions is satisfied:
\begin{eqnarray*}
(1)&&A_{1}=\pm A'_1, A_2 = \pm A'_2, A_3 = \pm A'_3, A_4 = \pm
 A'_4;
\\
(2)&&A_1 = \pm A'_1, A_2 = \pm A'_2, (A_3, A'_3 ) = (A_4, A'_4) =
 0;
\\
(3)&&A_1 = \pm A'_1, A_3 = \pm A'_3, (A_2, A'_2 ) = (A_4, A'_4) =
0;
\\
(4)&&A_1 = \pm A'_1, A_4 = \pm A'_4, (A_2, A'_2) = (A_3, A'_3) =
0;
\\
(5)&&A_2 = \pm A'_2, A_3 = \pm A'_3, (A_1, A'_1) = (A_4, A'_4 ) =
0;
\\
(6)&&A_2 = \pm A'_2, A_4 = \pm A'_4, (A_1, A'_1) = (A_3, A'_3) =
0;
\\
(7)&&A_3 = \pm A'_3, A_4 = \pm A'_4, (A_1, A'_1) = (A_2, A'_2) =
0;
\\
(8)&&(A_1, A'_1) = (A_2, A'_2) = (A_3, A'_3) = (A_4, A'_4) = 0.
\end{eqnarray*}This concludes that\begin{equation}
\{A_1 A_2 A_3 A_4, A'_1 A'_2 A_3 A_4, A_1 A_2 A'_3 A'_4 \},
\tag{2.4}\label{eq:2-4}\end{equation}\begin{equation}
\{A_{1}A_{2}A_{3}A_{4},A'_{1}A'_{2}A_{3}A_{4},A''_{1}A''_{2}A'_{3}A'_{4}\},
\tag{2.5}\label{eq:2-5}\end{equation}and\begin{equation}
\{A_{1}A_{2}A_{3}A_{4},A'_{1}A'_{2}A'_{3}A'_{4},A''_{1}A''_{2}A''_{3}A''_{4}\},
\tag{2.6}\label{eq:2-6}\end{equation} are all GHZ-Mermin
experiments of the four-qubit system, where
$$(A_j,A'_j)=(A_j,A''_j)=(A'_j,A''_j)=0$$for $j=1,2,3.$ In this
article, we need to clarify the geometric structure of GHZ-Mermin
experiments of the four-qubit system.

Browsing through the sets of mutually commuting four-qubit
nonlocal spin observables we quickly get the feeling that there
are many rather similar ones, and also some sets which can be
obtained in a rather trivial way (e.g., add a common element) from
2-qubit and 3-qubit ones. Hence, there are many equivalent
GHZ-Mermin experiments. Here, we describe the grouping of
GHZ-Mermin experiments into ``essentially distinct ones." Some
symmetries acting on GHZ-Mermin experiments are obvious. There are
two basic symmetries leading to equivalent experiments as follows.

$\mathrm{(S_1)}$ Changing the labelling of the local observables
at each site.

$\mathrm{(S_2)}$ Permuting systems.

Here, we define as equivalent two GHZ-Mermin experiments ${\cal
A}$ and ${\cal B}$ if they can be transformed to each other by
symmetrical actions $\mathrm{(S_1)}$ and $\mathrm{(S_2)}$ or local
unitary operations $\mathrm{(LU)}.$ In this case, we denote by
${\cal A} \cong {\cal B}.$ For example,
$$\{A''_1 A''_2 A_3 A_4, A'_1 A'_2 A_3 A_4, A''_1 A''_2 A'_3 A'_4
\} \cong \mathrm{Eq.(2.4)}$$by changing the labelling of the local
observables with $A_1 \longleftrightarrow A''_1$ and $A_2
\longleftrightarrow A''_2.$
Also,$$\{A_{1}A_{2}A_{3}A_{4},A'_{1}A_{2}A_{3}A'_{4},
A''_{1}A'_{2}A'_{3}A''_{4}\} \cong \mathrm{Eq.(2.5)}$$by permuting
the system with qubit 2 $\Longleftrightarrow$ qubit 4.

Since $\textrm{SU}(2) \cong \textrm{SO}(3)$ through $U^{\dagger}
(\vec{a} \vec{\sigma}) U = (R \vec{a}) \vec{\sigma},$ there is a
local unitary transformation $U_j$ on the $j$th qubit such that
$A_j = U^*_j \sigma^j_x U_j, A'_j = U^*_j \sigma^j_y U_j,$ and
$A''_j = U^*_j \sigma^j_z U_j,$ provided
$(A_j,A'_j)=(A_j,A''_j)=(A'_j,A''_j)=0.$ Then, the GHZ-Mermin
experiments Eqs.(2.4)-(2.6) are respectively equivalent
to\begin{equation} \{ \sigma^1_x \sigma^2_x \sigma^3_x \sigma^4_x,
\sigma^1_y \sigma^2_y \sigma^3_x \sigma^4_x, \sigma^1_x \sigma^2_x
\sigma^3_y \sigma^4_y \},
\tag{2.7}\label{eq:2-7}\end{equation}\begin{equation} \{
\sigma^1_x \sigma^2_x \sigma^3_x \sigma^4_x, \sigma^1_y \sigma^2_y
\sigma^3_x \sigma^4_x, \sigma^1_z \sigma^2_z \sigma^3_y \sigma^4_y
\}, \tag{2.8}\label{eq:2-8}\end{equation}and\begin{equation} \{
\sigma^1_x \sigma^2_x \sigma^3_x \sigma^4_x, \sigma^1_y \sigma^2_y
\sigma^3_y \sigma^4_y, \sigma^1_z \sigma^2_z \sigma^3_z \sigma^4_z
\}. \tag{2.9}\label{eq:2-9}
\end{equation}Moreover, since local observables are either
parallel or orthogonal to each other in a GHZ-Mermin experiment of
$n$ qubits by the above Lemma, it then must be equivalent to a
GHZ-Mermin experiment with each local observable taking one of
$\sigma_x, \sigma_y$ and $\sigma_z.$ Therefore, for constructing a
GHZ-Mermin experiment of $n$ qubits we only need to choose
$\sigma_x, \sigma_y,$ or $\sigma_z$ as local observables.

Clearly, Eqs.(2.7)-(2.9) possess different geometric structure.
That is, there are two dichotomic observables per site in
Eq.(2.7), two triads $(\sigma^1_x, \sigma^1_y, \sigma^1_z)$ and
$(\sigma^2_x, \sigma^2_y, \sigma^2_z)$ in Eq.(2.8), and four ones
in Eq.(2.9). Since symmetrical actions $\mathrm{(S_1)}$ and
$\mathrm{(S_2)}$ and local unitary operations $\mathrm{(LU)}$ do
not change the geometric structure of GHZ-Mermin experiments,
Eqs.(2.7)-(2.9) are inequivalent to each other. Generally
speaking, every GHZ-Mermin experiment has two geometric
invariants. On one hand, the number of sites which has a triad is
invariant under $\mathrm{(S_1)}, \mathrm{(S_2)},$ and
$\mathrm{(LU)},$ denoted by $\mathrm{C}.$ Clearly, $\mathrm{C}
\leq n$ for the $n$-qubit system. On the another hand, for every
element of the experiment there corresponds to the number of
elements which are orthogonal to that element at two sites. The
set of those numbers is also invariant under $\mathrm{(S_1)},
\mathrm{(S_2)},$ and $\mathrm{(LU)},$ denoted by $\mathrm{R}.$ For
example, the $\mathrm{C}$ and $\mathrm{R}$ invariants of Eq.(2.7)
are respectively $0$ and $(2, 1, 1),$ the ones of Eq.(2.8) are $2$
and $(1, 1, 0),$ and the ones of Eq.(2.9) are $4$ and $(0, 0, 0).$
Eqs.(2.7)-(2.9) have different geometric invariants. In the
sequel, we show that each four-qubit GHZ-Mermin experiment of
three elements must equivalently be one of the forms
Eqs.(2.7)-(2.9) and hence, two four-qubit GHZ-Mermin experiments
of three elements are equivalent if and only if they have the same
geometric invariants.

To this end, by $\mathrm{(S_1)}$ we have that each GHZ-Mermin
experiment of three elements must be one of the forms
\begin{equation} \{ \sigma^1_x \sigma^2_x \sigma^3_x \sigma^4_x,
\sigma^1_y \sigma^2_y \sigma^3_x \sigma^4_x, \star \star \star
\star \}, \tag{2.10}\label{eq:2-10}\end{equation}\begin{equation}
\{ \sigma^1_x \sigma^2_x \sigma^3_x \sigma^4_x, \sigma^1_y
\sigma^2_y \sigma^3_y \sigma^4_y, \star \star \star \star \},
\tag{2.11}\label{eq:2-11}\end{equation}because\begin{eqnarray*} \{
\sigma^1_x \sigma^2_x \sigma^3_x \sigma^4_x, \sigma^1_y \sigma^2_x
\sigma^3_y \sigma^4_x, \star \star \star \star \},
\\
\{ \sigma^1_x \sigma^2_x \sigma^3_x \sigma^4_x, \sigma^1_y
\sigma^2_x \sigma^3_x \sigma^4_y, \star \star \star \star \},
\\
\{ \sigma^1_x \sigma^2_x \sigma^3_x \sigma^4_x, \sigma^1_x
\sigma^2_y \sigma^3_y \sigma^4_x, \star \star \star \star \},
\\
\{ \sigma^1_x \sigma^2_x \sigma^3_x \sigma^4_x, \sigma^1_x
\sigma^2_y \sigma^3_x \sigma^4_y, \star \star \star \star \},
\\
\{ \sigma^1_x \sigma^2_x \sigma^3_x \sigma^4_x, \sigma^1_x
\sigma^2_x \sigma^3_y \sigma^4_y, \star \star \star \star \},
\end{eqnarray*}
are all equivalent to Eq.(2.10) by $\mathrm{(S_2)}.$ Since there
are at least two distinct observables at each site, Eq.(2.10)
reduces to Eq.(2.7), Eq.(2.8), and\begin{equation} \{ \sigma^1_x
\sigma^2_x \sigma^3_x \sigma^4_x, \sigma^1_y \sigma^2_y \sigma^3_x
\sigma^4_x, \sigma^1_y \sigma^2_y \sigma^3_y \sigma^4_y \}.
\tag{2.12}\label{eq:2-12}\end{equation}However, Eq.(2.12) is
equivalent to Eq.(2.7) by $\mathrm{(S_1)}$ with $\sigma^1_x
\longleftrightarrow \sigma^1_y$ and $\sigma^2_x
\longleftrightarrow \sigma^2_y.$

On the other hand, Eq.(2.11) reduces to Eq.(2.9),
\begin{eqnarray*}\{ \sigma^1_x \sigma^2_x \sigma^3_x
\sigma^4_x, \sigma^1_y \sigma^2_y \sigma^3_y \sigma^4_y,
\sigma^1_z \sigma^2_z \sigma^3_y \sigma^4_y \} \cong
\mathrm{Eq.(2.8)},\end{eqnarray*}and\begin{eqnarray*}\{ \sigma^1_x
\sigma^2_x \sigma^3_x \sigma^4_x, \sigma^1_y \sigma^2_y \sigma^3_y
\sigma^4_y, \sigma^1_z \sigma^2_z \sigma^3_x \sigma^4_x \} \cong
\mathrm{Eq.(2.8)}. \end{eqnarray*}This concludes the required
result.

\section{Trivial GHZ-Mermin Experiments}

There are many trivial GHZ-Mermin experiments of four qubits. For
example, GHZ-Mermin experiments Eqs.(2.7)-(2.9) are all trivial.
Indeed, we will show in this section that each of Eqs.(2.7)-(2.9)
is included in a GHZ-Mermin experiment of four elements and all
four-qubit GHZ-Mermin experiments of four elements are trivial.

In the following, we write $\sigma^1_x \sigma^2_x \sigma^3_y
\sigma^4_y,$ etc., as shorthand for $x x y y$ or $x_1 x_2 y_3
y_4.$ We characterize all four-qubit GHZ-Mermin experiments of
four elements as follows.

{\it Proposition: A $\mathrm{GHZ-Mermin}$ experiment of four
elements for the four-qubit system must equivalently be one of the
following forms:\begin{equation} \{ xxxx, yyxx, zzxx, xxyy\},
\tag{3.1}\label{eq:3-1}\end{equation}
\begin{equation} \{ xxxx, yyxx, yxyx, xxyy \},
\tag{3.2}\label{eq:3-2}\end{equation}\begin{equation} \{ xxxx,
yyxx, yxyx, yxxy \},
\tag{3.3}\label{eq:3-3}\end{equation}\begin{equation} \{ xxxx,
yyxx, yxyx, zzzy \},
\tag{3.4}\label{eq:3-4}\end{equation}\begin{equation} \{ xxxx,
yyxx, xxyy, yyyy \},
\tag{3.5}\label{eq:3-5}\end{equation}\begin{equation} \{ xxxx,
yyxx, xxyy, zzyy \},
\tag{3.6}\label{eq:3-6}\end{equation}\begin{equation}\{ xxxx,
yyxx, xxyy, zzzz \},
\tag{3.7}\label{eq:3-7}\end{equation}\begin{equation}\{ xxxx,
yyxx, zzyy, zzzz \}.
\tag{3.8}\label{eq:3-8}\end{equation}Moreover, the geometric
invariants of $\mathrm{Eqs.(3.1)-(3.8)}$ are illustrated in
$\mathrm{Table~ I}.$ }

{\it Proof}:~~At first, a GHZ-Mermin experiment of four elements
for the four-qubit system must equivalently be one of the
forms\begin{equation} \{ xxxx, yyxx, \star \star \star \star,
\star \star \star \star \},
\tag{3.9}\label{eq:3-9}\end{equation}\begin{equation} \{ xxxx,
yyyy, \star \star \star \star, \star \star \star \star \},
\tag{3.10}\label{eq:3-10}\end{equation}Then, Eq.(3.9) reduces to
one of the following forms\begin{equation} \{ xxxx, yyxx, zzxx,
\star \star \star \star \},
\tag{3.9-1}\label{eq:3-9-1}\end{equation}\begin{equation} \{ xxxx,
yyxx, yxyx, \star \star \star \star \},
\tag{3.9-2}\label{eq:3-9-2}\end{equation}\begin{equation} \{ xxxx,
yyxx, xxyy, \star \star \star \star \},
\tag{3.9-3}\label{eq:3-9-3}\end{equation}\begin{equation} \{ xxxx,
yyxx, zzyy, \star \star \star \star \},
\tag{3.9-4}\label{eq:3-9-4}\end{equation}because $\{ x x x x, y y
x x, y x x y, \star \star \star \star \},$ $\{ xxxx, yyxx, xyyx,
\star \star \star \star \},$ and $\{ xxxx, yyxx, xyxy, \star \star
\star \star \}$ are all equivalent to Eq.(3.9-2), as well as $\{
xxxx, yyxx, yyyy, \star \star \star \star \} \cong
\mathrm{Eq.(3.9-3)}.$

(1)~~From Eq.(3.9-1) we obtain Eq.(3.1) and\begin{eqnarray*}\{
xxxx, yyxx, zzxx, yyyy\} \cong \mathrm{Eq.(3.1)},
\\
\{ xxxx, yyxx, zzxx, zzyy\} \cong \mathrm{Eq.(3.1)}.
\end{eqnarray*}

(2)~~From Eq.(3.9-2) we obtain Eqs.(3.2)-(3.4)
and\begin{eqnarray*}\{ xxxx, yyxx, yxyx, xyxy \} \cong
\mathrm{Eq.(3.2)},
\\
\{ xxxx, yyxx, yxyx, yyyy \} \cong \mathrm{Eq.(3.2)}.
\end{eqnarray*}

(3)~~From Eq.(3.9-3) we obtain that Eqs.(3.1), (3.2), (3.5)-(3.7),
and\begin{eqnarray*}\{ xxxx, yyxx, xxyy, xxzz \} \cong
\mathrm{Eq.(3.1)},
\\
\{ xxxx, yyxx, xxyy, yxxy \} \cong \mathrm{Eq.(3.2)},
\\
\{ xxxx, yyxx, xxyy, xyyx \} \cong \mathrm{Eq.(3.2)},
\\
\{ xxxx, yyxx, xxyy, xyxy \} \cong \mathrm{Eq.(3.2)},
\\
\{ xxxx, yyxx, xxyy, yyzz \} \cong \mathrm{Eq.(3.6)}.
\end{eqnarray*}

(4)~~From Eq.(3.9-4) we obtain that Eqs.(3.6), (3.8),
and\begin{eqnarray*} \{ xxxx, yyxx, zzyy, zzxx \} \cong
\mathrm{Eq.(3.1)},
\\
\{ xxxx, yyxx, zzyy, yyyy \} \cong \mathrm{Eq.(3.6)},
\\
\{ xxxx, yyxx, zzyy, yyzz \} \cong \mathrm{Eq.(3.7)},
\\
\{ xxxx, yyxx, zzyy, xxzz \} \cong \mathrm{Eq.(3.7)}.
\end{eqnarray*}

On the other hand, since\begin{eqnarray*}\{ xxxx, yyyy, yyxx,
\star \star \star \star \},
\\
\{ xxxx, yyyy, zzxx, \star \star \star \star \},
\\
\{ xxxx, yyyy, zzyy, \star \star \star \star \},
\end{eqnarray*}and their variants are all included in Eq.(3.9), it
is concluded that Eq.(3.10) reduces to \begin{equation} \{ xxxx,
yyyy, zzzz, \star \star \star \star \},
\tag{3.10-1}\label{eq:3-10-1}\end{equation}.From Eq.(3.10-1) we
obtain\begin{eqnarray*}\{ xxxx, yyyy, zzzz, yyxx \} \cong
\mathrm{Eq.(3.7)},
\\
\{ xxxx, yyyy, zzzz, zzxx \} \cong
\mathrm{Eq.(3.7)},
\\
\{ xxxx, yyyy, zzzz, zzyy \} \cong
\mathrm{Eq.(3.7)}.\end{eqnarray*}
The proof is complete.

\begin{table}
\caption{\label{tab:table1}Here, we denote by $j (= 1,2,3,4)$ the
$j$-th element of the experiments. The numbers in the $\mathrm{C}$
column are $\mathrm{C}$-invariants, while the numbers in $1-4$'s
columns are $\mathrm{R}$-invariants.}
\begin{ruledtabular}
\begin{tabular}{cccccc}
 & $\mathrm{C}$ & 1 & 2 & 3 & 4 \\
\hline (3.1) & 2 & 3 & 2 & 2 & 1 \\
(3.2) & 0 & 3 & 2 & 3 & 2 \\
(3.3) & 0 & 3 & 3 & 3 & 3 \\
(3.4) & 3 & 2 & 2 & 2 & 0 \\
(3.5) & 0 & 2 & 2 & 2 & 2 \\
(3.6) & 2 & 2 & 1 & 2 & 1 \\
(3.7) & 4 & 2 & 1 & 1 & 0 \\
(3.8) & 4 & 1 & 1 & 1 & 1 \\
\end{tabular}
\end{ruledtabular}
\end{table}

From the above Proposition, it suffices to consider
Eqs.(3.1)-(3.8) for showing that every GHZ-Mermin experiments of
four elements for the four-qubit system is trivial. Let us recall
that the scenario for the GHZ-Mermin proof is the following:
Particles 1, 2, 3, and 4 move away from each other. At a given
time, an observer, Alice, has access to particle 1, a second
observer, Bob, has access to particle 2, a third observer,
Charlie, has access to particle 3, and a fourth observer, Davis,
has access to particle 4. For example, in the case of Eq.(3.5), by
introducing $(\cdot )$ to separate operators that can be viewed as
EPR's local elements of reality and for any common eigenstate
$|\varphi \rangle$ of Eq.(3.5), we have\begin{eqnarray*}
x_1 \cdot x_2 \cdot x_3 \cdot x_4 |\varphi \rangle & =
& \varepsilon_1 |\varphi \rangle,\\
y_1 \cdot y_2 \cdot x_3 \cdot x_4 |\varphi \rangle & =
& \varepsilon_2 |\varphi \rangle,\\
x_1 \cdot x_2 \cdot y_3 \cdot y_4 |\varphi \rangle & =
& \varepsilon_3 |\varphi \rangle,\\
y_1 \cdot y_2 \cdot y_3 \cdot y_4 |\varphi \rangle & = &
\varepsilon_4 |\varphi \rangle,
\end{eqnarray*}where $\varepsilon_j = \pm 1.$
According to EPR's criterion of local realism \cite{EPR}, Eq.(3.5)
allows Alice, Bob, Charlie, and Davis to predict the following
relations between the values of the elements of reality:
\begin{eqnarray*}
\nu( x_1 ) \nu( x_2 ) \nu( x_3 ) \nu( x_4 ) & = & \varepsilon_1,\\
\nu( y_1 ) \nu( y_2 ) \nu( x_3 ) \nu( x_4 ) & = & \varepsilon_2,\\
\nu( x_1 ) \nu( x_2 ) \nu( y_3 ) \nu( y_4 ) & = & \varepsilon_3,\\
\nu( y_1 ) \nu( y_2 ) \nu( y_3 ) \nu( y_4 ) & = & \varepsilon_4.
\end{eqnarray*}Since $ ( x_1 x_2 x_3 x_4 ) \times ( y_1 y_2 x_3 x_4) \times
( x_1 x_2 y_3 y_4 ) \times ( y_1 y_2 y_3 y_4 ) = 1,$ we have that
$\varepsilon_1 \varepsilon_2 \varepsilon_3 \varepsilon_4 = 1.$ In
this case, one can assign values $\nu( x_1 )= \varepsilon_1,$
$\nu( y_1 ) = \varepsilon_2, \nu( y_3 ) = \varepsilon_1
\varepsilon_3,$ and the remaining ones $\nu(\cdot) = 1.$ Thus, the
GHZ-Mermin proof is nullified in the case of Eq.(3.5).

The other cases are illustrated in Table II.

\begin{table}
\caption{\label{tab:table2} In every case, we can assign the
values as in the case of Eq.(3.5), the elements not indicated all
take the value $\nu(\cdot) = 1.$ }
\begin{ruledtabular}
\begin{tabular}{ccccc}
(3.1)& $\nu( x_1 ) = \varepsilon_1$ & $\nu( y_1 ) = \varepsilon_2$
& $ \nu( z_1 ) = \varepsilon_3$ &
$\nu( y_3 ) = \varepsilon_1 \varepsilon_4$\\
(3.2) & $\nu( x_1 ) = \varepsilon_1$ & $\nu( y_2 ) = \varepsilon_2
\varepsilon_3 $ & $\nu( y_1 ) = \varepsilon_3$ &
$\nu( y_4 ) = \varepsilon_1 \varepsilon_4 $\\
(3.3) & $\nu( x_1 ) = \varepsilon_1$ & $\nu( y_2 ) =
\varepsilon_2$ & $\nu( y_3 ) = \varepsilon_3$ &
$\nu( y_4 ) = \varepsilon_4$\\
(3.4) & $\nu( x_1 ) = \varepsilon_1$ & $\nu( y_2 ) =
\varepsilon_2$ & $\nu( y_3 ) = \varepsilon_3$ &
$\nu( z_1 ) = \varepsilon_4$\\
(3.6) & $\nu( x_1 ) = \varepsilon_1$ & $\nu( y_1 ) =
\varepsilon_2$ & $\nu( y_3 ) = \varepsilon_1 \varepsilon_3$ &
$\nu( z_1 ) = \varepsilon_1 \varepsilon_3 \varepsilon_4$\\
(3.7) & $\nu( x_1 ) = \varepsilon_1$ & $\nu( y_1 ) =
\varepsilon_2$ & $\nu( y_3 ) = \varepsilon_1 \varepsilon_3$ &
$\nu( z_1 ) = \varepsilon_4$\\
(3.8) & $\nu( x_1 ) = \varepsilon_1$ & $\nu( y_1 ) =
\varepsilon_2$ & $\nu( y_3 ) =  \varepsilon_3$ &
$\nu( z_3 ) = \varepsilon_4$
\end{tabular}
\end{ruledtabular}
\end{table}

Finally, we note that Eq.(2.7) is included in Eq.(3.1), Eq.(2.8)
in Eq.(3.6), and Eq.(2.9) (equivalently) in Eq.(3.7). This
concludes that the four-qubit GHZ-Mermin experiments of three
elements are all trivial. Therefore, a nontrivial GHZ-Mermin
experiment of four qubits must have at least five elements.

\section{Nontrivial GHZ-Mermin Experiments}

In this section, we will present a complete construction of
nontrivial four-qubit GHZ-Mermin experiments of five (6, 7, 8)
elements. We show that the experiments of five (6, 7, 8) elements
possess 11 (9, 5, 3) different forms. It is proved that in each
case there are two nontrivial GHZ-Mermin experiments and, the
associated states exhibiting an ``all versus nothing"
contradiction between quantum mechanics and $\mathrm{EPR}$'s local
realism are $\mathrm{GHZ}$ states.

\subsection{The case of five elements}

We first characterize all four-qubit GHZ-Mermin experiments of
five elements as follows.

{\it Proposition: A $\mathrm{GHZ-Mermin}$ experiment of five
elements for the four-qubit system must equivalently be one of the
following forms:\begin{equation} \{ xxxx, yyxx, zzxx, xxyy, xxzz
\}, \tag{4A.1}\label{eq:4A-1}\end{equation}\begin{equation} \{
xxxx, yyxx, zzxx, xxyy, yyyy \},
\tag{4A.2}\label{eq:4A-2}\end{equation}\begin{equation} \{ xxxx,
yyxx, zzxx, xxyy, yyzz \}, \tag{4A.3}\label{eq:4A-3}\end{equation}
\begin{equation} \{ xxxx, yyxx, yxyx, xxyy, yxxy \},
\tag{4A.4}\label{eq:4A-4}\end{equation}\begin{equation} \{ xxxx,
yyxx, yxyx, xxyy, xyxy \},
\tag{4A.5}\label{eq:4A-5}\end{equation}\begin{equation} \{ xxxx,
yyxx, yxyx, xxyy, zzzz \},
\tag{4A.6}\label{eq:4A-6}\end{equation}\begin{equation} \{ xxxx,
yyxx, yxyx, yxxy, zzzz \},
\tag{4A.7}\label{eq:4A-7}\end{equation}\begin{equation} \{ xxxx,
yyxx, yxyx, zzzy, xyyx \},
\tag{4A.8}\label{eq:4A-8}\end{equation}\begin{equation} \{ xxxx,
yyxx, xxyy, yyyy, zzzz \},
\tag{4A.9}\label{eq:4A-9}\end{equation}\begin{equation} \{ xxxx,
yyxx, xxyy, zzyy, yyzz \},
\tag{4A.10}\label{eq:4A-10}\end{equation}\begin{equation} \{ xxxx,
yyxx, xxyy, zzzz, yyzz \}.
\tag{4A.11}\label{eq:4A-11}\end{equation}Moreover, the geometric
invariants of $\mathrm{Eqs.(4A.1)-(4A.11)}$ are illustrated in
$\mathrm{Table~ III}.$ }

\begin{table}
\caption{\label{tab:table3}The numbers in the $\mathrm{C}$ line
are $\mathrm{C}$-invariants, while the numbers in $1-5$'s lines
are $\mathrm{R}$-invariants.}
\begin{ruledtabular}
\begin{tabular}{cccccccccccc}
 $\mathrm{4A}$& 1 & 2 & 3 & 4
 & 5 & 6 & 7 & 8 & 9 & 10 & 11\\
\hline $\mathrm{C}$ & 4 & 2 & 4 & 0 & 0 & 4 & 4 & 3 & 4 & 4 & 4 \\
\hline 1 & 4 & 3 & 3 & 4 & 4 & 3 & 3 & 3 & 2 & 2 & 2 \\
2 & 2 & 3 & 3 & 3 & 3 & 2 & 3 & 3 & 2 & 2 & 2 \\
3 & 2 & 2 & 2 & 4 & 3 & 3 & 3 & 3 & 2 & 2 & 1 \\
4 & 2 & 2 & 1 & 3 & 3 & 2 & 3 & 0 & 2 & 1 & 1 \\
5 & 2 & 2 & 1 & 4 & 3 & 0 & 0 & 3 & 0 & 1 & 2 \\
\end{tabular}
\end{ruledtabular}
\end{table}

{\it Proof}:~~By repeating the proof of the Proposition in Section
III, we find that every subset of four elements in a GHZ-Mermin
experiment of five elements for the four-qubit system is a
GHZ-Mermin experiment, i.e., a set of four mutually commuting
nonlocal spin observables with at least two different observables
at each site. By the Proposition in Sec.III, this concludes that a
four-qubit GHZ-Mermin experiment of five elements must
equivalently be one of the forms\begin{equation} \{ xxxx, yyxx,
zzxx, xxyy, \star \star \star \star \},
\tag{4A.12}\label{eq:4A-12}\end{equation}
\begin{equation} \{ xxxx, yyxx, yxyx, xxyy, \star \star \star \star \},
\tag{4A.13}\label{eq:4A-13}\end{equation}\begin{equation} \{ xxxx,
yyxx, yxyx, yxxy, \star \star \star \star \},
\tag{4A.14}\label{eq:4A-14}\end{equation}\begin{equation} \{ xxxx,
yyxx, yxyx, zzzy, \star \star \star \star \},
\tag{4A.15}\label{eq:4A-15}\end{equation}\begin{equation} \{ xxxx,
yyxx, xxyy, yyyy, \star \star \star \star \},
\tag{4A.16}\label{eq:4A-16}\end{equation}\begin{equation} \{ xxxx,
yyxx, xxyy, zzyy, \star \star \star \star \},
\tag{4A.17}\label{eq:4A-17}\end{equation}\begin{equation} \{ xxxx,
yyxx, xxyy, zzzz, \star \star \star \star \},
\tag{4A.18}\label{eq:4A-18}\end{equation}\begin{equation} \{ xxxx,
yyxx, zzyy, zzzz, \star \star \star \star \}.
\tag{4A.19}\label{eq:4A-19}\end{equation}

(1)~~From Eq.(4A.12), we obtain Eqs.(4A.1)-(4A.3),
and\begin{eqnarray*}\{ xxxx, yyxx, zzxx, xxyy, zzyy \} \cong
\mathrm{Eq.(4A.2)},\\[0.3cm]
\{ xxxx, yyxx, zzxx, xxyy, zzzz \} \cong \mathrm{Eq.(4A.3)}.
\end{eqnarray*}

(2)~~From Eq.(4A.13), we obtain Eqs.(4A.4)-(4A.6),
and\begin{eqnarray*}\{ xxxx, yyxx, yxyx, xxyy, xyyx \} \cong
\mathrm{Eq.(4A.4)},\\[0.3cm]
\{ xxxx, yyxx, yxyx, xxyy, yyyy \} \cong \mathrm{Eq.(4A.5)}.
\end{eqnarray*}

(3)~~From Eq.(4A.14), we obtain Eqs.(4A.4), (4A.7),
and\begin{eqnarray*}\{ xxxx, yyxx, yxyx, yxxy, xyyx \} \cong
\mathrm{Eq.(4A.4)},\\[0.3cm]
\{ xxxx, yyxx, yxyx, yxxy, xyxy \} \cong
\mathrm{Eq.(4A.4)},\\[0.3cm]
\{ xxxx, yyxx, yxyx, yxxy, yyyy \} \cong \mathrm{Eq.(4A.4)}.
\end{eqnarray*}

(4)~~From Eq.(4A.15), we obtain Eq.(4A.8) and\begin{eqnarray*}\{
xxxx, yyxx,yxyx, zzzy, yyyz \} \cong
\mathrm{Eq.(4A.6)},\\[0.3cm]
\{ xxxx, yyxx, yxyx, zzzy, yxxz \} \cong
\mathrm{Eq.(4A.7)},\\[0.3cm]
\{ xxxx, yyxx, yxyx, zzzy, xyxz \} \cong
\mathrm{Eq.(4A.6)},\\[0.3cm]
\{ xxxx, yyxx, yxyx, zzzy, xxyz \} \cong \mathrm{Eq.(4A.6)}.
\end{eqnarray*}

(5)~~From Eq.(4A.16), we obtain Eqs.(4A.2), (4A.9),
and\begin{eqnarray*} \{ xxxx, yyxx, xxyy, yyyy, yxyx \} \cong
\mathrm{Eq.(4A.5)},\\[0.3cm]
\{ xxxx, yyxx, xxyy, yyyy, yxxy \} \cong
\mathrm{Eq.(4A.5)},\\[0.3cm]
\{ xxxx, yyxx, xxyy, yyyy, xyyx \} \cong
\mathrm{Eq.(4A.5)},\\[0.3cm]
\{ xxxx, yyxx, xxyy, yyyy, xyxy \} \cong
\mathrm{Eq.(4A.5)},\\[0.3cm]
\{ xxxx, yyxx, xxyy, yyyy, xxzz \} \cong
\mathrm{Eq.(4A.2)},\\[0.3cm]
\{ xxxx, yyxx, xxyy, yyyy, yyzz \} \cong
\mathrm{Eq.(4A.2)},\\[0.3cm]
\{ xxxx, yyxx, xxyy, yyyy, zzyy \} \cong \mathrm{Eq.(4A.2)}.
\end{eqnarray*}

(6)~~From Eq.(4A.17), we obtain Eq.(4A.10) and\begin{eqnarray*}\{
xxxx, yyxx, xxyy, zzyy, zzxx \} \cong
\mathrm{Eq.(4A.2)}, \\[0.3cm]
\{ xxxx, yyxx, xxyy, zzyy, xxzz \} \cong
\mathrm{Eq.(4A.3)}, \\[0.3cm]
\{ xxxx, yyxx, xxyy, zzyy, yyyy \} \cong
\mathrm{Eq.(4A.2)}, \\[0.3cm]
\{ xxxx, yyxx, xxyy, zzyy, zzzz \} \cong \mathrm{Eq.(4A.10)}.
\end{eqnarray*}

(7)~~From Eq.(4A.18), we obtain Eqs.(4A.6), (4A.9), (4A.11),
and\begin{eqnarray*}\{ xxxx, yyxx, xxyy, zzzz, yxxy \} \cong
\mathrm{Eq.(4A.6)}, \\[0.3cm]
\{ xxxx, yyxx, xxyy, zzzz, xyyx \} \cong
\mathrm{Eq.(4A.6)}, \\[0.3cm]
\{ xxxx, yyxx, xxyy, zzzz, xyxy \} \cong
\mathrm{Eq.(4A.6)}, \\[0.3cm]
\{ xxxx, yyxx, xxyy, zzzz, zzxx \} \cong
\mathrm{Eq.(4A.3)}, \\[0.3cm]
\{ xxxx, yyxx, xxyy, zzzz, zzyy \} \cong \mathrm{Eq.(4A.11)},
\\[0.3cm]
\{ xxxx, yyxx, xxyy, zzzz, xxzz \} \cong
\mathrm{Eq.(4A.3)}.
\end{eqnarray*}

(8)~~From Eq.(4A.19), we obtain
\begin{eqnarray*}\{ xxxx, yyxx, zzyy, zzzz, zzxx \} \cong
\mathrm{Eq.(4A.1)}, \\[0.3cm]
\{ xxxx, yyxx, zzyy, zzzz, xxyy \} \cong
\mathrm{Eq.(4A.11)}, \\[0.3cm]
\{ xxxx, yyxx, zzyy, zzzz, xxzz \} \cong
\mathrm{Eq.(4A.11)}, \\[0.3cm]
\{ xxxx, yyxx, zzyy, zzzz, yyzz \} \cong
\mathrm{Eq.(4A.11)}, \\[0.3cm]
\{ xxxx, yyxx, xxyy, zzzz, yyyy \} \cong
\mathrm{Eq.(4A.11)}.
\end{eqnarray*}

From Table III we find that except for Eqs.(4A.10) and (4A.11),
each of Eqs.(4A.1)-(4A.11) has different geometric invariants and
hence, they are inequivalent. In order to distinguish Eq.(4A.10)
from Eq.(4A.11), we need to use other geometric invariants. Note
that, for every subset of three elements in a GHZ-Mermin
experiment there corresponds the number of sites at which there is
a triad. Those numbers are invariant under $\mathrm{(S_1)}$ and
$\mathrm{(S_2)}.$ For example, $\{xxxx, yyzz, zzyy\}$ in
Eq.(4A.10) has four triads, while $\{xxxx, zzzz, yyzz\}$ in
Eq.(4A.11) has two triads. It is evident that there is no subset
of three elements in Eq.(4A.11) possessing four triads. This
concludes that Eqs.(4A.10) and (4A.11) are inequivalent. The proof
is complete.

\begin{table}
\caption{\label{tab:table4} The elements not indicated all take
the value $\nu(\cdot) = 1.$ }
\begin{ruledtabular}
\begin{tabular}{cccc}
(4A.1)& $\nu( x_1 ) = \varepsilon_1$ & $\nu( y_1 ) =
\varepsilon_2$ & $ \nu( z_1 ) = \varepsilon_3$\\
& $\nu( y_3 ) = \varepsilon_1 \varepsilon_4$ &
$\nu( z_3 ) = \varepsilon_1 \varepsilon_5 $ &\\
\hline (4A.3) & $\nu( x_1 ) = \varepsilon_1$ & $\nu( y_1 ) =
\varepsilon_2 $ & $\nu( z_1 ) = \varepsilon_3$\\
& $\nu( y_3 ) = \varepsilon_1 \varepsilon_4 $ &
$\nu( z_3 ) = \varepsilon_2 \varepsilon_5$ & \\
\hline (4A.6) & $\nu( x_1 ) = \varepsilon_1$ & $\nu( y_2 ) =
\varepsilon_2$ & $\nu( y_3 ) = \varepsilon_3$\\
 & $\nu( y_4 ) = \varepsilon_1 \varepsilon_3 \varepsilon_4$ &
$\nu( z_1 ) = \varepsilon_5$ & \\
\hline (4A.7) & $\nu( x_1 ) = \varepsilon_1$ & $\nu( y_2 ) =
\varepsilon_2$ & $\nu( y_3 ) = \varepsilon_3$ \\
& $\nu( y_4 ) = \varepsilon_4$ & $\nu( z_1 ) = \varepsilon_5$ &\\
\hline (4A.10) & $\nu( x_1 ) = \varepsilon_1$ & $\nu( y_1 ) =
\varepsilon_2$ & $\nu( y_3 ) = \varepsilon_1 \varepsilon_3$ \\
& $\nu( z_1 ) = \varepsilon_1 \varepsilon_3 \varepsilon_4$ &
$\nu( z_3 ) = \varepsilon_2 \varepsilon_5$ &\\
\hline (4A.11) & $\nu( x_1 ) = \varepsilon_1$ & $\nu( y_1 ) =
\varepsilon_2$ & $\nu( y_3 ) = \varepsilon_1 \varepsilon_3$ \\
& $\nu( z_1 ) = \varepsilon_2 \varepsilon_4 \varepsilon_5$ & $\nu(
z_3 ) = \varepsilon_2 \varepsilon_5$ &
\end{tabular}
\end{ruledtabular}
\end{table}

It is easy to see that Eqs.(4A.1), (4A.3), (4A.6), (4A.7),
(4A.10), and (4A.11) are trivial, whose assigned values are
illustrated in Table IV. As follows, we show that Eqs.(4A.2),
(4A.5), and (4A.9) are also trivial. Indeed, we note that $(x_1
x_2 x_3 x_4) \times (y_1 y_2 x_3 x_4) \times (x_1 x_2 y_3 y_4)
\times (y_1 y_2 y_3 y_4) = 1.$ This concludes that for Eq.(4A.2),
$\varepsilon_1 \varepsilon_2 \varepsilon_4 \varepsilon_5 = 1$ and
thus, one can assign $\nu (x_1) = \varepsilon_1, \nu (y_1) =
\varepsilon_2, \nu (z_1) = \varepsilon_3, \nu (y_3) =
\varepsilon_1 \varepsilon_4,$ and the remaining ones $\nu ( \cdot
) =1.$ Similarly, since $(y_1 y_2 x_3 x_4) \times (y_1 x_2 y_3
x_4) \times (x_1 y_2 x_3 y_4) \times (x_1 x_2 y_3 y_4) = 1$ in
Eq.(4A.5) and $(x_1 x_2 x_3 x_4) \times (y_1 y_2 x_3 x_4) \times
(x_1 x_2 y_3 y_4) \times (y_1 y_2 y_3 y_4) = 1$ in Eq.(4A.9)
respectively, it is easily concluded that Eqs.(4A.5) and (4A.9)
are both trivial.

In the sequel, we prove that both Eqs.(4A.4) and (4A.8) are
nontrivial, and the associated states exhibiting 100\% violation
between quantum mechanics and EPR's local realism are GHZ states.

{\it Theorem: Nontrivial $\mathrm{GHZ-Mermin}$ experiments of five
elements for the four-qubit system must equivalently be either
$\mathrm{Eq.(4A.4)}$ or $\mathrm{Eq.(4A.8)}.$ Moreover, the
associated states exhibiting an ``all versus nothing"
contradiction between quantum mechanics and $\mathrm{EPR}$'s local
realism are $\mathrm{GHZ}$ states.}

{\it Proof}.~~At first, for $| \varphi \rangle =
\frac{1}{\sqrt{2}} \left ( |0000 \rangle - |1111 \rangle \right )$
we have\begin{equation} x_1 x_2 x_3 x_4 |\varphi \rangle = - |
\varphi \rangle,
\tag{4A.19}\label{eq:4A-19}\end{equation}\begin{equation} y_1 y_2
x_3 x_4 | \varphi \rangle = | \varphi \rangle,
\tag{4A.20}\label{eq:4A-20}\end{equation}\begin{equation} y_1 x_2
y_3 x_4 | \varphi \rangle = | \varphi \rangle,
\tag{4A.21}\label{eq:4A-21}\end{equation}\begin{equation} x_1 x_2
y_3 y_4 | \varphi \rangle = | \varphi \rangle,
\tag{4A.22}\label{eq:4A-22}\end{equation}\begin{equation} y_1 x_2
x_3 y_4 | \varphi \rangle = | \varphi \rangle.
\tag{4A.23}\label{eq:4A-23}\end{equation}By the GHZ-Mermin
argument based on EPR's local realism, one has\begin{equation} \nu
(x_1) \nu (x_2) \nu (x_3) \nu (x_4) = - 1,
\tag{4A.24}\label{eq:4A-24}\end{equation}\begin{equation} \nu
(y_1) \nu (y_2) \nu (x_3) \nu (x_4) = 1,
\tag{4A.25}\label{eq:4A-25}\end{equation}\begin{equation} \nu
(y_1) \nu (x_2) \nu (y_3) \nu (x_4) = 1,
\tag{4A.26}\label{eq:4A-26}\end{equation}\begin{equation} \nu
(x_1) \nu (x_2) \nu (y_3) \nu (y_4) = 1,
\tag{4A.27}\label{eq:4A-27}\end{equation}\begin{equation} \nu
(y_1) \nu (x_2) \nu (x_3) \nu (y_4) = 1.
\tag{4A.28}\label{eq:4A-28}\end{equation}However,
Eqs.(4A.24)-(4A.28) are inconsistent, because when we take the
product of Eqs.(4A.24) and (4A.26)-(4A.28), the value of the
left-hand side is one, while the right-hand side is $-1.$ This
concludes that Eq.(4A.4) is nontrivial.

Although the inconsistence of Eqs.(4A.24)-(4A.28) is concluded
from Eqs.(4A.24) and (4A.26)-(4A.28), the subset of $\{ xxxx,
yxyx, xxyy, yxxy\}$ in Eq.(4A.4) is not a GHZ-Mermin experiment of
four qubits at whose second site there is only one measurement. On
the other hand, there are some states other than GHZ's states
satisfying Eqs.(4A.19) and (4A.21)-(4A.23), such as $| \psi
\rangle = a \left ( | 0000 \rangle - | 1111 \rangle \right ) + b
\left ( | 0100 \rangle - | 1011 \rangle \right )$ with $|a |^2 +
|b |^2 = 1/2.$ However, we will show that the states exhibiting
the GHZ-Mermin proof in Eq.(4A.4) are the GHZ states. Thus,
Eq.(4A.20) and so $yyxx$ plays a crucial role in the GHZ-Mermin
experiment Eq.(4A.4).

Similarly, we can prove that Eq.(4A.8) is also nontrivial and omit
the details. In the sequel, we prove that the GHZ state is the
unique state with equivalence up to a local unitary transformation
which presents the GHZ-Mermin proof in both Eqs.(4A.4) and (4A.8).

To this end, we consider the generic form of Eq.(4A.8) and,
suppose $|\varphi \rangle$ is the common eigenstate of five
commuting nonlocal spin observables such that
\begin{equation}
A_1 A_2 A_3 A_4 |\varphi\rangle = \varepsilon_1 |\varphi\rangle,
\tag{4A.29}\label{eq:4A-29}\end{equation}\begin{equation} A'_1
A'_2 A_3 A_4 |\varphi\rangle = \varepsilon_2 |\varphi\rangle,
\tag{4A.30}\label{eq:4A-30}\end{equation}\begin{equation} A'_1 A_2
A'_3 A_4 |\varphi\rangle = \varepsilon_3 |\varphi\rangle,
\tag{4A.31}\label{eq:4A-31}\end{equation}\begin{equation} A_1 A'_2
A'_3 A_4 |\varphi\rangle = \varepsilon_4 |\varphi\rangle,
\tag{4A.32}\label{eq:4A-32}\end{equation}\begin{equation} A''_1
A''_2 A''_3 A'_4 |\varphi\rangle = \varepsilon_5 |\varphi\rangle,
\tag{4A.33}\label{eq:4A-33}\end{equation}where $(A, A') = (A, A'')
= (A', A'') =0.$ According to GHZ-Mermin's analysis based on EPR's
local realism, it is concluded that
\begin{equation}
\nu (A_1 ) \nu (A_2) \nu (A_3) \nu (A_4) =
\varepsilon_1,\tag{4A.34}\label{eq:4A-34}\end{equation}\begin{equation}
\nu(A'_1) \nu(A'_2) \nu(A_3) \nu(A_4) = \varepsilon_2,
\tag{4A.35}\label{eq:4A-35}\end{equation}\begin{equation}
\nu(A'_1) \nu(A_2) \nu(A'_3) \nu(A_4) = \varepsilon_3,
\tag{4A.36}\label{eq:4A-36}\end{equation}\begin{equation} \nu(A_1)
\nu(A'_2) \nu(A'_3) \nu(A_4) = \varepsilon_4,
\tag{4A.37}\label{eq:4A-37}\end{equation}\begin{equation}
\nu(A''_1) \nu(A''_2) \nu(A''_3) \nu(A'_4) = \varepsilon_5.
\tag{4A.38}\label{eq:4A-38}\end{equation}When $\varepsilon_1
\varepsilon_2 \varepsilon_3 \varepsilon_4 = 1,$ one can assign
$\nu(A_1) = \varepsilon_1, \nu(A'_2) = \varepsilon_2, \nu(A'_3) =
\varepsilon_3, \nu(A'_4) = \varepsilon_5,$ and the remaining ones
$\nu(\cdot)=1.$ Therefore, the necessary condition for $|\varphi
\rangle$ presenting a GHZ-Mermin-type proof is
\begin{equation}
\varepsilon_1 \varepsilon_2 \varepsilon_3 \varepsilon_4 = - 1.
\tag{4A.39}\label{eq:4A-39}\end{equation}

On the other hand, suppose Eq.(4A.39) holds, then it is impossible
to assign values, either 1 or -1, that satisfy Eqs.(4A.35)-(4A.38)
because when take the product of Eqs.(4A.35)-(4A.38), the value of
the left hand is equal to $\varepsilon_5$ while the right hand is
$-\varepsilon_5.$ Thus the condition Eq.(4A.39) is the necessary
and sufficient condition for $|\varphi\rangle$ presenting a
GHZ-Mermin-type proof. In this case, by changing the signs of
local observables $A_j$ and $A'_j$ ($A_1 \rightarrow -
\varepsilon_1 A_1, A'_2 \rightarrow \varepsilon_2 A'_2, A'_3
\rightarrow \varepsilon_3 A'_3,$ and $A'_4 \rightarrow
\varepsilon_5 A'_4$), we have that
\begin{equation}
A_1 A_2 A_3 A_4 |\varphi \rangle = -|\varphi \rangle,
\tag{4A.40}\label{eq:4A-40}\end{equation}\begin{equation} A'_1
A'_2 A_3 A_4 |\varphi \rangle = |\varphi \rangle,
\tag{4A.41}\label{eq:4A-41}\end{equation}\begin{equation} A'_1 A_2
A'_3 A_4 |\varphi \rangle = |\varphi \rangle,
\tag{4A.42}\label{eq:4A-42}\end{equation}\begin{equation} A_1 A'_2
A'_3 A_4  |\varphi \rangle = |\varphi \rangle,
\tag{4A.43}\label{eq:4A-43}\end{equation}\begin{equation} A''_1
A''_2 A''_3 A'_4 |\varphi \rangle = |\varphi\rangle.
\tag{4A.44}\label{eq:4A-44}\end{equation}

By Eqs.(2.1)-(2.3), one has that
\begin{eqnarray*}
A_j A'_j = -A'_j A_j = i A''_j,\\[0.3cm]
A'_j A''_j = - A''_j A'_j = i A_j,\\[0.3cm]
A''_j A_j = -A_j A''_j = i A'_j,\\[0.3cm]
A^2_j = (A'_j )^2 = (A''_j )^2 = 1.
\end{eqnarray*}
Hence, $A_j, A'_j,$ and $A''_j$ satisfy the algebraic identities
of Pauli's matrices \cite{Pauli}. Therefore, choosing $A''_j$
representation $\{|0 \rangle_j,|1 \rangle_j \},$ i.e., $ A''_j |0
\rangle_j = |0 \rangle_j, A''_j |1 \rangle_j = - |1 \rangle_j,$ we
have that
\begin{eqnarray*}A_j |0 \rangle_j = e^{-i\alpha_j} |1 \rangle_j,
~~A_j |1 \rangle_j = e^{i\alpha_j} |0 \rangle_j,\\[0.3cm]
A'_j |0 \rangle_j =i e^{-i\alpha_j} |1 \rangle_j,~~ A'_j |1
\rangle_j = -ie^{i\alpha_j} |0 \rangle_j,
\end{eqnarray*}where $0 \leq \alpha_j \leq 2 \pi.$ We write
$|0100 \rangle,$ etc., as shorthand for $|0 \rangle_1 \otimes |1
\rangle_2 \otimes |0 \rangle_3 \otimes |0 \rangle_4.$ Since
$\{|\epsilon_1 \epsilon_2 \epsilon_3 \epsilon_4 \rangle :
\epsilon_1, \epsilon_2, \epsilon_3, \epsilon_4 =0,1 \}$ is an
orthogonal basis of the four-qubit system. We can uniquely write:
\begin{eqnarray*}
|\varphi \rangle = \sum_{\epsilon_1, \epsilon_2, \epsilon_3,
\epsilon_4 =0,1} \lambda_{\epsilon_1 \epsilon_2 \epsilon_3
\epsilon_4 }|\epsilon_1 \epsilon_2 \epsilon_3 \epsilon_4 \rangle
\end{eqnarray*}
with $\sum|\lambda_{\epsilon_1 \epsilon_2 \epsilon_3
\epsilon_4}|^2 = 1.$ We define the four-qubit operator
\begin{eqnarray*}
 \mathcal{B} = -A_1 A_2 A_3 A_4 + A'_1 A'_2 A_3 A_4\\
 + A'_1 A_2 A'_3 A_4+ A_1 A'_2 A'_3 A_4.
\end{eqnarray*}
Then by Eqs.(4A.40)-(4A.43), one has that $\mathcal{B} |\varphi
\rangle = 4|\varphi \rangle.$ This conclude that
\begin{equation}
 \mathcal{B}^2 |\varphi \rangle = 16 |\varphi \rangle.
\tag{4A.45}\label{eq:4A-45}\end{equation}However, a simple
computation yields that
\begin{eqnarray*}
\mathcal{B}^2 = 4 + 4 (A''_1 A''_2 + A''_1 A''_3 + A''_2 A''_3 ).
\end{eqnarray*}
Then, by using Eq.(4A.45) we conclude that
\begin{eqnarray*}
|\varphi \rangle = a |0000 \rangle + b |0001 \rangle + c |1110
\rangle + d |1111 \rangle
\end{eqnarray*}where $a = \lambda_{0000 }, b = \lambda_{0001},
c = \lambda_{1110},$ and $d = \lambda_{1111}.$ From Eqs.(4A.40)
and (4A.44) it is concluded that $a = -d e^{i ( \alpha_1 + \alpha_2
+ \alpha_3 + \alpha_4) }, b = -id e^{i ( \alpha_1 + \alpha_2 +
\alpha_3 )},$ and $ c = i d e^{i \alpha_4}.$ Therefore
\begin{eqnarray*}
|\varphi \rangle = -d e^{i ( \alpha_1 + \alpha_2 + \alpha_3 )}(
e^{i \alpha_4} |0000 \rangle + i |0001 \rangle ) \\
+ d ( i e^{i
\alpha_4} |1110 \rangle + |1111 \rangle )\\
= \frac{1}{\sqrt{2}} ( e^{i \theta} | 000 \rangle | u \rangle +
e^{i \phi}| 111 \rangle | v \rangle )
\end{eqnarray*}where $ | u \rangle = \frac{1}{\sqrt{2}} ( e^{i \alpha_4} | 0 \rangle + i
| 1 \rangle ), | v \rangle =  \frac{1}{\sqrt{2}} ( i e^{i
\alpha_4} | 0 \rangle + | 1 \rangle),$ and $0 \leq \theta, \phi
\leq 2 \pi.$ Since $\langle u | v \rangle =0,$ $|\varphi \rangle$
is a GHZ state.

The proof for Eq.(4A.4) is similar and omitted.

\subsection{The case of six elements}

We first characterize all four-qubit GHZ-Mermin experiments of six
elements as follows.

{\it Proposition: A $\mathrm{GHZ-Mermin}$ experiment of six
elements for the four-qubit system must equivalently be one of the
following forms:\begin{equation} \{ xxxx, yyxx, zzxx, xxyy, xxzz,
yyyy \}, \tag{4B.1}\label{eq:4B-1}\end{equation}\begin{equation}
\{ xxxx, yyxx, zzxx, xxyy, yyyy, zzyy \},
\tag{4B.2}\label{eq:4B-2}\end{equation}\begin{equation} \{ xxxx,
yyxx, zzxx, xxyy, yyyy, zzzz \},
\tag{4B.3}\label{eq:4B-3}\end{equation}\begin{equation} \{ xxxx,
yyxx, zzxx, xxyy, yyzz, zzzz \},
\tag{4B.4}\label{eq:4B-4}\end{equation}
\begin{equation} \{ xxxx, yyxx, yxyx, xxyy, yxxy, xyyx \},
\tag{4B.5}\label{eq:4B-5}\end{equation}\begin{equation} \{ xxxx,
yyxx, yxyx, xxyy, yxxy, zzzz \},
\tag{4B.6}\label{eq:4B-6}\end{equation}\begin{equation} \{ xxxx,
yyxx, yxyx, xxyy, xyxy, yyyy \},
\tag{4B.7}\label{eq:4B-7}\end{equation}\begin{equation} \{ xxxx,
yyxx, yxyx, xxyy, xyxy, zzzz \},
\tag{4B.8}\label{eq:4B-8}\end{equation}\begin{equation} \{ xxxx,
yyxx, xxyy, zzyy, yyzz, zzzz \}.
\tag{4B.9}\label{eq:4B-9}\end{equation}Moreover, the geometric
invariants of $\mathrm{Eqs.(4B.1)-(4B.9)}$ are illustrated in
$\mathrm{Table~ V}.$ }

\begin{table}
\caption{\label{tab:table5}The numbers in the $\mathrm{C}$ line
are $\mathrm{C}$-invariants, while the numbers in $1-6$'s lines
are $\mathrm{R}$-invariants.}
\begin{ruledtabular}
\begin{tabular}{cccccccccc}
 $\mathrm{4B}$& 1 & 2 & 3 & 4
 & 5 & 6 & 7 & 8 & 9\\
\hline $\mathrm{C}$ & 4 & 2 & 4 & 4 & 0 & 4 & 0 & 4 & 4\\
\hline 1 & 4 & 3 & 3 & 3 & 5 & 4 & 4 & 4 & 2 \\
2 & 3 & 3 & 3 & 3 & 4 & 3 & 4 & 3 & 2 \\
3 & 2 & 3 & 3 & 3 & 5 & 4 & 4 & 3 & 2 \\
4 & 3 & 3 & 2 & 1 & 4 & 3 & 4 & 3 & 2 \\
5 & 2 & 3 & 2 & 2 & 4 & 4 & 4 & 3 & 2 \\
6 & 2 & 3 & 1 & 2 & 4 & 0 & 4 & 0 & 2 \\
\end{tabular}
\end{ruledtabular}
\end{table}

{\it Proof}:~~By the same argument in the case of five elements,
it is concluded that a subset of five elements in a GHZ-Mermin
experiment of six elements for the four-qubit system is a
four-qubit GHZ-Mermin experiment of five elements. Then, by the
Proposition in Section IV.A, a four-qubit GHZ-Mermin experiment of
six elements must equivalently be one of the forms\begin{equation}
\{ xxxx, yyxx, zzxx, xxyy, xxzz, \star \star \star \star \},
\tag{4B.10}\label{eq:4B-10}\end{equation}\begin{equation} \{ xxxx,
yyxx, zzxx, xxyy, yyyy, \star \star \star \star \},
\tag{4B.11}\label{eq:4B-11}\end{equation}\begin{equation} \{ xxxx,
yyxx, zzxx, xxyy, yyzz, \star \star \star \star \},
\tag{4B.12}\label{eq:4B-12}\end{equation}\begin{equation} \{ xxxx,
yyxx, yxyx, xxyy, yxxy, \star \star \star \star \},
\tag{4B.13}\label{eq:4B-13}\end{equation}\begin{equation} \{ xxxx,
yyxx, yxyx, xxyy, xyxy, \star \star \star \star \},
\tag{4B.14}\label{eq:4B-14}\end{equation}\begin{equation} \{ xxxx,
yyxx, yxyx, xxyy, zzzz, \star \star \star \star \},
\tag{4B.15}\label{eq:4B-15}\end{equation}\begin{equation} \{ xxxx,
yyxx, yxyx, yxxy, zzzz, \star \star \star \star \},
\tag{4B.16}\label{eq:4B-16}\end{equation}\begin{equation} \{ xxxx,
yyxx, yxyx, zzzy, xyyx, \star \star \star \star \},
\tag{4B.17}\label{eq:4B-17}\end{equation}\begin{equation} \{ xxxx,
yyxx, xxyy, yyyy, zzzz, \star \star \star \star \},
\tag{4B.18}\label{eq:4B-18}\end{equation}\begin{equation} \{ xxxx,
yyxx, xxyy, zzyy, yyzz, \star \star \star \star \},
\tag{4B.19}\label{eq:4B-19}\end{equation}\begin{equation} \{ xxxx,
yyxx, xxyy, zzzz, yyzz, \star \star \star \star \}.
\tag{4B.20}\label{eq:4B-20}\end{equation}

(1)~~From Eq.(4B.10) we obtain Eq.(4B.1), and\begin{eqnarray*} \{
xxxx, yyxx, zzxx, xxyy, xxzz, yyzz \} \cong
\mathrm{Eq.(4B.1)}, \\[0.3cm] \{ xxxx, yyxx, zzxx, xxyy,
xxzz, zzyy \} \cong
\mathrm{Eq.(4B.1)}, \\[0.3cm]
\{ xxxx, yyxx, zzxx, xxyy, xxzz, zzzz \} \cong
\mathrm{Eq.(4B.1)}.\end{eqnarray*}

(2)~~From Eq.(4B.11) we obtain Eqs.(4B.2), (4B.3),
and\begin{eqnarray*}\{ xxxx, yyxx, zzxx, xxyy, yyyy, xxzz \} \cong
\mathrm{Eq.(4B.1)}, \\[0.3cm]
\{ xxxx, yyxx, zzxx, xxyy, yyyy, yyzz \} \cong
\mathrm{Eq.(4B.1)}.\end{eqnarray*}

(3)~~From Eq.(4B.12) we obtain Eq.(4B.4), and\begin{eqnarray*}\{
xxxx, yyxx, zzxx, xxyy, yyzz, xxzz \} \cong
\mathrm{Eq.(4B.1)}, \\[0.3cm]
\{ xxxx, yyxx, zzxx, xxyy, yyzz, yyyy \} \cong
\mathrm{Eq.(4B.1)}, \\[0.3cm]
\{ xxxx, yyxx, zzxx, xxyy, yyzz, zzyy \} \cong
\mathrm{Eq.(4B.3)}.\end{eqnarray*}

(4)~~From Eq.(4B.13) we obtain Eqs.(4B.5), (4B.6),
and\begin{eqnarray*} \{ xxxx, yyxx, yxyx, xxyy, yxxy, xyxy \}
\cong \mathrm{Eq.(4B.5)}, \\[0.3cm]
\{ xxxx, yyxx, yxyx, xxyy, yxxy, yyyy \} \cong
\mathrm{Eq.(4B.5)}.\end{eqnarray*}

(5)~~From Eq.(4B.14) we obtain Eqs.(4B.7), (4B.8),
and\begin{eqnarray*} \{ xxxx, yyxx, yxyx, xxyy, xyxy, yxxy \}
\cong \mathrm{Eq.(4B.5)}, \\[0.3cm]
\{ xxxx, yyxx, yxyx, xxyy, xyxy, xyyx \} \cong
\mathrm{Eq.(4B.5)}.\end{eqnarray*}

(6)~~From Eq.(4B.15) we obtain Eqs.(4B.6), (4B.8),
and\begin{eqnarray*} \{ xxxx, yyxx, yxyx, xxyy, zzzz, xyyx \}
\cong \mathrm{Eq.(4B.6)}, \\[0.3cm]
\{ xxxx, yyxx, yxyx, xxyy, zzzz, yyyy \} \cong
\mathrm{Eq.(4B.8)}.\end{eqnarray*}

(7)~~From Eq.(4B.16) we obtain Eq.(4B.6) and\begin{eqnarray*} \{
xxxx, yyxx, yxyx, yxxy, zzzz, xyyx \}
\cong \mathrm{Eq.(4B.6)}, \\[0.3cm]
\{ xxxx, yyxx, yxyx, yxxy, zzzz, xyxy \}
\cong \mathrm{Eq.(4B.6)}, \\[0.3cm]
\{ xxxx, yyxx, yxyx, yxxy, zzzz, yyyy \} \cong
\mathrm{Eq.(4B.6)}.\end{eqnarray*}

(8)~~From Eq.(4B.17) we obtain\begin{eqnarray*} \{ xxxx, yyxx,
yxyx, zzzy, xyyx, yyyz \} \cong \mathrm{Eq.(4B.6)},\\[0.3cm]
\{ xxxx, yyxx, yxyx, zzzy, xyyx, yxxz \}
\cong \mathrm{Eq.(4B.6)},\\[0.3cm]
\{ xxxx, yyxx, yxyx, zzzy, xyyx, xyxz \}
\cong \mathrm{Eq.(4B.6)},
\\[0.3cm]
\{ xxxx, yyxx, yxyx, zzzy, xyyx, xxyz \}
\cong \mathrm{Eq.(4B.6)}.
\end{eqnarray*}

(9)~~From Eq.(4B.18) we obtain Eq.(4B.3) and\begin{eqnarray*} \{
xxxx, yyxx, xxyy, yyyy, zzzz, yxyx \}
\cong \mathrm{Eq.(4B.8)}, \\[0.3cm]
\{ xxxx, yyxx, xxyy, yyyy, zzzz, yxxy \}
\cong \mathrm{Eq.(4B.8)}, \\[0.3cm]
\{ xxxx, yyxx, xxyy, yyyy, zzzz, xyyx \}
\cong \mathrm{Eq.(4B.8)}, \\[0.3cm]
\{ xxxx, yyxx, xxyy, yyyy, zzzz, xyxy \}
\cong \mathrm{Eq.(4B.8)}, \\[0.3cm]
\{ xxxx, yyxx, xxyy, yyyy, zzzz, xxzz \}
\cong \mathrm{Eq.(4B.3)}, \\[0.3cm]
\{ xxxx, yyxx, xxyy, yyyy, zzzz, yyzz \}
\cong \mathrm{Eq.(4B.3)}, \\[0.3cm]
\{ xxxx, yyxx, xxyy, yyyy, zzzz, zzyy \} \cong
\mathrm{Eq.(4B.3)}.\end{eqnarray*}

(10)~~From Eq.(4B.19) we obtain Eq.(4B.9) and\begin{eqnarray*} \{
xxxx, yyxx, xxyy, zzyy, yyzz, zzxx \} \cong \mathrm{Eq.(4B.3)},
\\[0.3cm]
\{ xxxx, yyxx, xxyy, zzyy, yyzz, xxzz \} \cong \mathrm{Eq.(4B.3)},
\\[0.3cm]
\{ xxxx, yyxx, xxyy, zzyy, yyzz, yyyy \} \cong \mathrm{Eq.(4B.1)},
\\[0.3cm] \{ xxxx, yyxx, xxyy,
zzyy, yyzz, xxzz \} \cong \mathrm{Eq.(4B.3)}.\end{eqnarray*}

(11)~~From Eq.(4B.20) we obtain Eqs.(4B.4), (4B.9)
and\begin{eqnarray*} \{ xxxx, yyxx, xxyy, zzzz, yyzz, xxzz \}
\cong \mathrm{Eq.(4B.1)},\\[0.3cm]
\{ xxxx, yyxx, xxyy, zzzz, yyzz, yyyy \} \cong
\mathrm{Eq.(4B.3)}.\end{eqnarray*}

From Table V we find that except for Eqs.(4B.3) and (4B.4), each
of Eqs.(4A.1)-(4A.9) has different geometric invariants and hence,
they are inequivalent. However, $\{xxxx, yyzz, zzyy\}$ in
Eq.(4B.3) has four triads, while there is no subset of three
elements in Eq.(4B.4) possessing four triads, as noted in the case
of Eqs.(4A.10) and (4A.11) this concludes that Eqs.(4B.3) and
(4B.4) are inequivalent. The proof is complete.

{\it Theorem: Nontrivial $\mathrm{GHZ-Mermin}$ experiments of six
elements for the four-qubit system must equivalently be either
$\mathrm{Eq.(4B.5)}$ or $\mathrm{Eq.(4B.6)}.$ Moreover, the
associated states exhibiting an ``all versus nothing"
contradiction between quantum mechanics and $\mathrm{EPR}$'s local
realism are $\mathrm{GHZ}$ states.}

{\it Proof}.~~By the above Proposition, it suffices to show that
Eqs.(4B.1)-(4B.4) and (4B.7)-(4B.9) are all trivial, while
Eqs.(4B.5) and (4B.6) are both nontrivial.

(1)~~Since $(xxxx)\times (yyxx) \times (xxyy) \times (yyyy) = 1,$
for Eq.(4B.1) we have $\varepsilon_1 \varepsilon_2 \varepsilon_4
\varepsilon_6 =1.$ Then, one can assign $\nu (x_1) =
\varepsilon_1, \nu (y_1) = \varepsilon_2, \nu (z_1) =
\varepsilon_3, \nu (y_3) = \varepsilon_1 \varepsilon_4, \nu (z_3)
= \varepsilon_1 \varepsilon_5,$ and the remaining ones $v( \cdot )
=1.$

(2)~~Since $(xxxx)\times (yyxx) \times (xxyy) \times (yyyy) = 1$
and $(xxxx)\times (zzxx) \times (xxyy) \times (zzyy) = 1,$ for
Eq.(4B.2) we have $\varepsilon_1 \varepsilon_2 \varepsilon_4
\varepsilon_5 =1$ and $\varepsilon_1 \varepsilon_3 \varepsilon_4
\varepsilon_6 =1,$ respectively. Then, one can assign $\nu (x_1) =
\varepsilon_1, \nu (y_1) = \varepsilon_2, \nu (z_1) =
\varepsilon_3, \nu (y_3) = \varepsilon_1 \varepsilon_4,$ and the
remaining ones $v( \cdot ) =1.$

(3)~~Since $(xxxx)\times (yyxx) \times (xxyy) \times (yyyy) = 1,$
for Eq.(4B.3) we have $\varepsilon_1 \varepsilon_2 \varepsilon_4
\varepsilon_5 = 1.$ Then, one can assign $\nu (x_1) =
\varepsilon_1, \nu (y_1) = \varepsilon_2, \nu (z_1) =
\varepsilon_3, \nu (y_3) = \varepsilon_1 \varepsilon_4, \nu (z_3)
= \varepsilon_3 \varepsilon_6,$ and the remaining ones $v( \cdot )
=1.$

(4)~~Since $(yyxx)\times (zzxx) \times (yyzz) \times (zzzz) = 1,$
for Eq.(4B.4) we have $\varepsilon_2 \varepsilon_3 \varepsilon_5
\varepsilon_6 = 1.$ Then, one can assign $\nu (x_1) =
\varepsilon_1, \nu (y_1) = \varepsilon_2, \nu (z_1) =
\varepsilon_3, \nu (y_3) = \varepsilon_1 \varepsilon_4, \nu (z_3)
= \varepsilon_2 \varepsilon_5,$ and the remaining ones $v( \cdot )
=1.$

(5)~~Since $(yyxx)\times (yxyx) \times (xxyy) \times (xyxy) = 1$
and $(xxxx)\times (yxyx) \times (xyxy) \times (yyyy) = 1,$ for
Eq.(4B.7) we have $\varepsilon_2 \varepsilon_3 \varepsilon_4
\varepsilon_5 =1$ and $\varepsilon_1 \varepsilon_3 \varepsilon_5
\varepsilon_6 =1,$ respectively. Then, one can assign $\nu (x_1) =
\varepsilon_1, \nu (y_2) = \varepsilon_2, \nu (y_3) =
\varepsilon_3, \nu (y_4) = \varepsilon_1 \varepsilon_3
\varepsilon_4,$ and the remaining ones $v( \cdot ) =1.$

(6)~~Since $(yyxx)\times (yxyx) \times (xxyy) \times (xyxy) = 1,$
for Eq.(4B.8) we have $\varepsilon_2 \varepsilon_3 \varepsilon_4
\varepsilon_5 = 1.$ Then, one can assign $\nu (x_1) =
\varepsilon_1, \nu (y_2) = \varepsilon_2, \nu (y_3) =
\varepsilon_3, \nu (y_4) = \varepsilon_1 \varepsilon_3
\varepsilon_4, \nu (z_1) = \varepsilon_6,$ and the remaining ones
$v( \cdot ) =1.$

(7)~~Since $(xxxx) \times (yyxx) \times (xxyy) \times (zzxx)
\times (yyzz) \times (zzzz) = 1,$ for Eq.(4B.9) we have
$\varepsilon_1 \varepsilon_2 \varepsilon_3 \varepsilon_4
\varepsilon_5 \varepsilon_6 = 1.$ Then, one can assign $\nu (x_1)
= \varepsilon_1, \nu (y_1) = \varepsilon_2, \nu (y_3) =
\varepsilon_1 \varepsilon_3, \nu (z_1) = \varepsilon_1
\varepsilon_3 \varepsilon_4, \nu (z_3) = \varepsilon_2
\varepsilon_5,$ and the remaining ones $v( \cdot ) =1.$

Since Eq.(4A.4) is included in Eqs.(4B.5) and (4B.6) and $|
\varphi \rangle = \frac{1}{\sqrt{2}} \left ( |0000 \rangle - |1111
\rangle \right )$ is a common eigenstate of both Eqs.(4B.5) and
(4B.6), it is concluded that Eqs.(4B.5) and (4B.6) are both
nontrivial. Moreover, as shown in Section IV.A that the GHZ state
is the unique state with equivalence up to a local unitary
transformation which presents the GHZ-Mermin proof in Eq.(4A.4),
we conclude the same result for Eqs.(4B.5) and (4B.6). This
completes the proof.

\subsection{The case of seven elements}

We characterize all four-qubit GHZ-Mermin experiments of seven
elements as follows.

{\it Proposition: A $\mathrm{GHZ-Mermin}$ experiment of seven
elements for the four-qubit system must equivalently be one of the
following forms:\begin{equation} \{ xxxx, yyxx, zzxx, xxyy, xxzz,
yyyy, zzzz \}, \tag{4C.1}\label{eq:4C-1}
\end{equation}\begin{equation} \{ xxxx, yyxx, zzxx, xxyy, xxzz,
yyyy, zzyy \}, \tag{4C.2}\label{eq:4C-2}
\end{equation}\begin{equation} \{ xxxx, yyxx, yxyx, xxyy, yxxy,
xyyx, xyxy \}, \tag{4C.3}\label{eq:4C-3}
\end{equation}\begin{equation} \{ xxxx, yyxx, yxyx, xxyy, yxxy,
xyyx, zzzz \}, \tag{4C.4}\label{eq:4C-4}
\end{equation}\begin{equation} \{ xxxx, yyxx, yxyx, xxyy, xyxy,
yyyy, zzzz \}. \tag{4C.5}\label{eq:4C-5}
\end{equation}Moreover, the geometric
invariants of $\mathrm{Eqs.(4C.1)-(4C.5)}$ are illustrated in
$\mathrm{Table~ VI}.$ }

\begin{table}
\caption{\label{tab:table6}The numbers in the $\mathrm{C}$ column
are $\mathrm{C}$-invariants, while the numbers in $1-7$'s columns
are $\mathrm{R}$-invariants.}
\begin{ruledtabular}
\begin{tabular}{ccccccccc}
 & $\mathrm{C}$ & 1 & 2 & 3 & 4 & 5 & 6 & 7\\
\hline (4C.1) & 4 & 4 & 3 & 3 & 3 & 3 & 2 & 2 \\
(4C.2) & 4 & 4 & 3 & 3 & 4 & 2 & 3 & 3 \\
(4C.3) & 0 & 6 & 5 & 5 & 5 & 5 & 5 & 5 \\
(4C.4) & 4 & 5 & 4 & 5 & 4 & 4 & 4 & 0 \\
(4C.5) & 4 & 4 & 4 & 4 & 4 & 4 & 4 & 0 \\
\end{tabular}
\end{ruledtabular}
\end{table}

{\it Proof}:~~As similar as above, a subset of six elements in a
GHZ-Mermin experiment of seven elements for the four-qubit system
is a four-qubit GHZ-Mermin experiment of six elements. Then, by
the Proposition in Section IV.B, a four-qubit GHZ-Mermin
experiment of seven elements must equivalently be one of the
forms\begin{equation} \{ xxxx, yyxx, zzxx, xxyy, xxzz, yyyy, \star
\star \star \star \},
\tag{4C.6}\label{eq:4C-6}\end{equation}\begin{equation} \{ xxxx,
yyxx, zzxx, xxyy, yyyy, zzyy, \star \star \star \star \},
\tag{4C.7}\label{eq:4C-7}\end{equation}\begin{equation} \{ xxxx,
yyxx, zzxx, xxyy, yyyy, zzzz, \star \star \star \star \},
\tag{4C.8}\label{eq:4C-8}\end{equation}\begin{equation} \{ xxxx,
yyxx, zzxx, xxyy, yyzz, zzzz, \star \star \star \star \},
\tag{4C.9}\label{eq:4C-9}\end{equation}\begin{equation} \{ xxxx,
yyxx, yxyx, xxyy, yxxy, xyyx, \star \star \star \star \},
\tag{4C.10}\label{eq:4C-10}\end{equation}\begin{equation} \{ xxxx,
yyxx, yxyx, xxyy, yxxy, zzzz, \star \star \star \star \},
\tag{4C.11}\label{eq:4C-11}\end{equation}\begin{equation} \{ xxxx,
yyxx, yxyx, xxyy, xyxy, yyyy, \star \star \star \star \},
\tag{4C.12}\label{eq:4C-12}\end{equation}\begin{equation} \{ xxxx,
yyxx, yxyx, xxyy, xyxy, zzzz, \star \star \star \star \},
\tag{4C.13}\label{eq:4C-13}\end{equation}\begin{equation} \{ xxxx,
yyxx, xxyy, zzyy, yyzz, zzzz, \star \star \star \star \}.
\tag{4C.14}\label{eq:4C-14}\end{equation}

(1)~~From Eq.(4C.6) we obtain Eqs.(4C.1), (4C.2),
and\begin{eqnarray*} \{ xxxx, yyxx, zzxx, xxyy, xxzz, yyyy, yyzz
\} \cong \mathrm{Eq.(4C.2)}.\end{eqnarray*}

(2)~~From Eq.(4C.7) we obtain Eq.(4C.2) and\begin{eqnarray*} \{
xxxx, yyxx, zzxx, xxyy, yyyy, zzyy, zzzz \}\cong
\mathrm{Eq.(4C.2)}, \\[0.3cm]
\{ xxxx, yyxx, zzxx, xxyy, yyyy, zzyy, yyzz \}\cong
\mathrm{Eq.(4C.2)}.\end{eqnarray*}

(3)~~From Eq.(4C.8) we obtain Eq.(4C.1) and\begin{eqnarray*} \{
xxxx, yyxx, zzxx, xxyy, yyyy, zzzz, yyzz \} \cong
\mathrm{Eq.(4C.1)}, \\[0.3cm]
\{ xxxx, yyxx, zzxx, xxyy, yyyy, zzzz, zzyy \} \cong
\mathrm{Eq.(4C.2)}.\end{eqnarray*}

(4)~~From Eq.(4C.9) we obtain\begin{eqnarray*}\{ xxxx, yyxx, zzxx,
xxyy, yyzz, zzzz, xxzz \} \cong \mathrm{Eq.(4C.2)}, \\[0.3cm]
\{ xxxx, yyxx, zzxx, xxyy, yyzz, zzzz, yyyy \} \cong
\mathrm{Eq.(4C.1)}, \\[0.3cm]
\{ xxxx, yyxx, zzxx, xxyy, yyzz, zzzz, zzyy \} \cong
\mathrm{Eq.(4C.1)}.\end{eqnarray*}

(5)~~From Eq.(4C.10) we obtain Eqs.(4C.3), (4C.4),
and\begin{eqnarray*} \{ xxxx, yyxx, yxyx, xxyy, yxxy, xyyx, yyyy
\} \cong \mathrm{Eq.(4C.3)}.\end{eqnarray*}

(6)~~From Eq.(4C.11) we obtain Eq.(4C.4) and\begin{eqnarray*}\{
xxxx, yyxx, yxyx, xxyy, yxxy, zzzz, xyxy \}
\cong \mathrm{Eq.(4C.4)}, \\[0.3cm]
\{ xxxx, yyxx, yxyx, xxyy, yxxy, zzzz, yyyy \} \cong
\mathrm{Eq.(4C.4)}.\end{eqnarray*}

(7)~~From Eq.(4C.12) we obtain Eq.(4C.5) and\begin{eqnarray*}\{
xxxx, yyxx, yxyx, xxyy, xyxy, yyyy, yxxy \}
\cong \mathrm{Eq.(4C.3)}, \\[0.3cm]
\{ xxxx, yyxx, yxyx, xxyy, xyxy, yyyy, xyyx \} \cong
\mathrm{Eq.(4C.3)}.\end{eqnarray*}

(8)~~From Eq.(4C.13) we obtain Eq.(4C.5), and\begin{eqnarray*} \{
xxxx, yyxx, yxyx, xxyy, xyxy, zzzz, yxxy
\} \cong \mathrm{Eq.(4C.4)}, \\[0.3cm]
\{ xxxx, yyxx, yxyx, xxyy, xyxy, zzzz, xyyx \} \cong
\mathrm{Eq.(4C.4)}.
\end{eqnarray*}

(9)~~From Eq.(4C.14) we obtain\begin{eqnarray*}\{ xxxx, yyxx,
xxyy, zzyy, yyzz, zzzz, zzxx \} \cong
\mathrm{Eq.(4C.1)}, \\[0.3cm]
\{ xxxx, yyxx, xxyy, zzyy, yyzz, zzzz, xxzz \} \cong
\mathrm{Eq.(4C.1)}, \\[0.3cm]
\{ xxxx, yyxx, xxyy, zzyy, yyzz, zzzz, yyyy \} \cong
\mathrm{Eq.(4C.1)}.\end{eqnarray*}

From Table VI we find that each of Eqs.(4C.1)-(4C.5) has different
geometric invariants and hence, they are all inequivalent. The
proof is complete.

{\it Theorem: Nontrivial $\mathrm{GHZ-Mermin}$ experiments of
seven elements for the four-qubit system must equivalently be
either $\mathrm{Eq.(4C.3)}$ or $\mathrm{Eq.(4C.4)}.$ Moreover, the
associated states exhibiting an ``all versus nothing"
contradiction between quantum mechanics and $\mathrm{EPR}$'s local
realism are $\mathrm{GHZ}$ states.}

{\it Proof}.~~By the above Proposition, it suffices to show that
Eqs.(4C.1), (4C.2), and (4C.5) are all trivial, while Eqs.(4C.3)
and (4C.4) are both nontrivial.

(1)~~Since $(xxxx)\times (yyxx) \times (xxyy) \times (yyyy) = 1$
and $(xxxx)\times (zzxx) \times (xxzz) \times (zzzz) = 1,$ for
Eq.(4C.1) we have $\varepsilon_1 \varepsilon_2 \varepsilon_4
\varepsilon_6 =1$ and $\varepsilon_1 \varepsilon_3 \varepsilon_5
\varepsilon_7 =1,$ respectively. Then, one can assign $\nu (x_1) =
\varepsilon_1, \nu (y_1) = \varepsilon_2, \nu (z_1) =
\varepsilon_3, \nu (y_3) = \varepsilon_1 \varepsilon_4, \nu (z_3)
= \varepsilon_1 \varepsilon_5,$ and the remaining ones $v( \cdot )
=1.$

(2)~~Since $(xxxx)\times (yyxx) \times (xxyy) \times (yyyy) = 1$
and $(xxxx)\times (zzxx) \times (xxyy) \times (zzyy) = 1,$ for
Eq.(4C.2) we have $\varepsilon_1 \varepsilon_2 \varepsilon_4
\varepsilon_6 =1$ and $\varepsilon_1 \varepsilon_3 \varepsilon_4
\varepsilon_7 =1,$ respectively. Then, one can assign $\nu (x_1) =
\varepsilon_1, \nu (y_1) = \varepsilon_2, \nu (z_1) =
\varepsilon_3, \nu (y_3) = \varepsilon_1 \varepsilon_4, \nu (z_3)
= \varepsilon_1 \varepsilon_5,$ and the remaining ones $v( \cdot )
=1.$

(3)~~Since $(xxxx)\times (yxyx) \times (xyxy) \times (yyyy) = 1$
and $(yyxx)\times (yxyx) \times (xxyy) \times (xyxy) = 1,$ for
Eq.(4C.5) we have $\varepsilon_1 \varepsilon_3 \varepsilon_5
\varepsilon_6 =1$ and $\varepsilon_2 \varepsilon_3 \varepsilon_4
\varepsilon_5 =1,$ respectively. Then, one can assign $\nu (x_1) =
\varepsilon_1, \nu (y_2) = \varepsilon_2, \nu (y_3) =
\varepsilon_3, \nu (y_4) = \varepsilon_1 \varepsilon_3
\varepsilon_4, \nu (z_1) = \varepsilon_7,$ and the remaining ones
$v( \cdot ) =1.$

Since Eq.(4B.5) is included in Eqs.(4C.3) and (4C.4) and $|
\varphi \rangle = \frac{1}{\sqrt{2}} \left ( |0000 \rangle - |1111
\rangle \right )$ is a common eigenstate of both Eqs.(4C.3) and
(4B.4), it is concluded that Eqs.(4C.3) and (4C.4) are both
nontrivial. Moreover, as shown in Section IV.B that the GHZ state
is the unique state with equivalence up to a local unitary
transformation which presents the GHZ-Mermin proof in Eq.(4B.5),
we conclude the same result for Eqs.(4C.3) and (4C.4). This
completes the proof.

\subsection{The case of eight elements}

We characterize all four-qubit GHZ-Mermin experiments of eight
elements as follows.

{\it Proposition: A $\mathrm{GHZ-Mermin}$ experiment of eight
elements for the four-qubit system must equivalently be one of the
following forms:\begin{equation} \{ xxxx, yyxx, zzxx, xxyy, xxzz,
yyyy, zzzz, yyzz \}, \tag{4D.1}\label{eq:4D-1}
\end{equation}\begin{equation} \{ xxxx, yyxx, yxyx, xxyy, yxxy,
xyyx, xyxy, yyyy \}, \tag{4D.2}\label{eq:4D-2}
\end{equation}\begin{equation} \{ xxxx, yyxx, yxyx, xxyy, yxxy,
xyyx, xyxy, zzzz \}. \tag{4D.3}\label{eq:4D-3}
\end{equation}Moreover, the geometric
invariants of $\mathrm{Eqs.(4D.1)-(4D.3)}$ are illustrated in
$\mathrm{Table~ VII}.$ }

\begin{table}
\caption{\label{tab:table7}The numbers in the $\mathrm{C}$ column
are $\mathrm{C}$-invariants, while the numbers in $1-8$'s columns
are $\mathrm{R}$-invariants.}
\begin{ruledtabular}
\begin{tabular}{cccccccccc}
 & $\mathrm{C}$ & 1 & 2 & 3 & 4 & 5 & 6 & 7 & 8\\
\hline (4D.1) & 4 & 4 & 4 & 3 & 2 & 4 & 3 & 3 & 4 \\
(4D.2) & 0 & 6 & 6 & 6 & 6 & 6 & 6 & 6 & 6 \\
(4D.3) & 4 & 6 & 5 & 5 & 5 & 5 & 5 & 5 & 0 \\
\end{tabular}
\end{ruledtabular}
\end{table}

{\it Proof}:~~As similar as above, a subset of seven elements in a
GHZ-Mermin experiment of eight elements for the four-qubit system
is a four-qubit GHZ-Mermin experiment of seven elements. Then, by
the Proposition in Section IV.C, a four-qubit GHZ-Mermin
experiment of eight elements must equivalently be one of the
forms\begin{equation} \{ xxxx, yyxx, zzxx, xxyy, xxzz, yyyy, zzzz,
\star \star \star \star \}, \tag{4D.4}\label{eq:4D-4}
\end{equation}\begin{equation} \{ xxxx, yyxx, zzxx, xxyy, xxzz,
yyyy, zzyy, \star \star \star \star \}, \tag{4D.5}\label{eq:4D-5}
\end{equation}\begin{equation} \{ xxxx, yyxx, yxyx, xxyy, yxxy,
xyyx, xyxy, \star \star \star \star \}, \tag{4D.6}\label{eq:4D-6}
\end{equation}\begin{equation} \{ xxxx, yyxx, yxyx, xxyy, yxxy,
xyyx, zzzz, \star \star \star \star \}, \tag{4D.7}\label{eq:4D-7}
\end{equation}\begin{equation} \{ xxxx, yyxx, yxyx, xxyy, xyxy,
yyyy, zzzz, \star \star \star \star \}. \tag{4D.8}\label{eq:4D-8}
\end{equation}

(1)~~From Eq.(4D.4) we obtain Eq.(4D.1) and\begin{eqnarray*}\{
xxxx, yyxx, zzxx, xxyy, xxzz, yyyy, zzzz, zzyy \}\\[0.3cm] \cong
\mathrm{Eq.(4D.1)}.\end{eqnarray*}

(2)~~From Eq.(4D.5) we obtain\begin{eqnarray*}\{ xxxx, yyxx, zzxx,
xxyy, xxzz,  yyyy, zzyy, zzzz \} \\[0.3cm] \cong
\mathrm{Eq.(4D.1)},\\[0.3cm]
\{ xxxx, yyxx, zzxx, xxyy, xxzz,  yyyy, zzyy, yyzz \} \\[0.3cm] \cong
\mathrm{Eq.(4D.1)} .\end{eqnarray*}

(3)~~From Eq.(4D.6) we obtain Eqs.(4D.2) and (4D.3).

(4)~~From Eq.(4D.7) we obtain Eq.(4D.3) and\begin{eqnarray*}\{
xxxx, yyxx, yxyx, xxyy, yxxy, xyyx, zzzz, yyyy \} \\[0.3cm] \cong
\mathrm{Eq.(4D.3)}.\end{eqnarray*}

(5)~~From Eq.(4D.8) we obtain\begin{eqnarray*}\{ xxxx, yyxx, yxyx,
xxyy, xyxy,  yyyy, zzzz, yxxy \} \\[0.3cm] \cong
\mathrm{Eq.(4D.3)},\\[0.3cm]
\{ xxxx, yyxx, yxyx, xxyy, xyxy, yyyy, zzzz, xyyx \} \\[0.3cm] \cong
\mathrm{Eq.(4D.3)}.\end{eqnarray*}

From Table VII we find that each of Eqs.(4D.1)-(4D.3) has
different geometric invariants and hence, they are all
inequivalent. The proof is complete.

{\it Theorem: Nontrivial $\mathrm{GHZ-Mermin}$ experiments of
eight elements for the four-qubit system must equivalently be
either $\mathrm{Eq.(4D.2)}$ or $\mathrm{Eq.(4D.3)}.$ Moreover, the
associated states exhibiting an ``all versus nothing"
contradiction between quantum mechanics and $\mathrm{EPR}$'s local
realism are $\mathrm{GHZ}$ states.}

{\it Proof}.~~By the above Proposition, it suffices to show that
Eq.(4D.1) is trivial, while Eqs.(4D.2) and (4D.3) are both
nontrivial.

Indeed, since $(xxxx)\times (yyxx) \times (xxyy) \times (yyyy) =
1, (xxxx)\times (zzxx) \times (xxzz) \times (zzzz) = 1, $ and
$(xxxx)\times (yyxx) \times (xxzz) \times (yyzz) = 1,$ for
Eq.(4D.1) we have $\varepsilon_1 \varepsilon_2 \varepsilon_4
\varepsilon_6 =1, \varepsilon_1 \varepsilon_3 \varepsilon_5
\varepsilon_7 =1,$ and $\varepsilon_1 \varepsilon_2 \varepsilon_5
\varepsilon_8 =1.$ Then, one can assign $\nu (x_1) =
\varepsilon_1, \nu (y_1) = \varepsilon_2, \nu (z_1) =
\varepsilon_3, \nu (y_3) = \varepsilon_1 \varepsilon_4, \nu (z_3)
= \varepsilon_1 \varepsilon_5,$ and the remaining ones $v( \cdot )
=1.$

On the other hand, Eq.(4C.3) is included in Eqs.(4D.2) and (4D.3)
and $| \varphi \rangle = \frac{1}{\sqrt{2}} \left ( |0000 \rangle
- |1111 \rangle \right )$ is a common eigenstate of both
Eqs.(4D.2) and (4D.3), it is concluded that Eqs.(4D.2) and (4D.3)
are both nontrivial. Moreover, as shown in Section IV.C that the
GHZ state is the unique state with equivalence up to a local
unitary transformation which presents the GHZ-Mermin proof in
Eq.(4C.3), we conclude the same result for Eqs.(4D.2) and (4D.3).
This completes the proof.

\section{Maximal GHZ-Mermin Experiments}

In this section, we show that a GHZ-Mermin experiment of the
four-qubit system contains at most nine elements and those maximal
GHZ-Mermin experiments of nine elements have two different forms,
one of which is trivial, while another one is nontrivial.

{\it Proposition: A $\mathrm{GHZ-Mermin}$ experiment of the
four-qubit system contains at most nine elements, and a four-qubit
$\mathrm{GHZ-Mermin}$ experiment of nine elements must
equivalently be one of the following forms:\begin{equation} \{
xxxx, yyxx, zzxx, xxyy, xxzz, yyyy, zzzz, yyzz, zzyy \},
\tag{5.1}\label{eq:5-1}
\end{equation}\begin{equation} \{ xxxx, yyxx, yxyx, xxyy, yxxy,
xyyx, xyxy, yyyy, zzzz \}. \tag{5.2}\label{eq:5-2}
\end{equation}Moreover, the geometric
invariants of $\mathrm{Eqs.(5.1)}$ and $\mathrm{(5.2)}$ are
illustrated in $\mathrm{Table~ VIII}.$ }

\begin{table}
\caption{\label{tab:table8}The numbers in the $\mathrm{C}$ column
are $\mathrm{C}$-invariants, while the numbers in $1-9$'s columns
are $\mathrm{R}$-invariants.}
\begin{ruledtabular}
\begin{tabular}{ccccccccccc}
 & $\mathrm{C}$ & 1 & 2 & 3 & 4 & 5 & 6 & 7 & 8 & 9\\
\hline (5.1) & 4 & 4 & 4 & 4 & 4 & 4 & 4 & 4 & 4 & 4 \\
(5.2) & 4 & 6 & 6 & 6 & 6 & 6 & 6 & 6 & 6 & 0 \\
\end{tabular}
\end{ruledtabular}
\end{table}

{\it Proof}:~~Indeed, as similar as above, a subset of eight
elements in a GHZ-Mermin experiment of nine elements for the
four-qubit system is a four-qubit GHZ-Mermin experiment of eight
elements. Then, by the Proposition in Section IV.D, a four-qubit
GHZ-Mermin experiment of nine elements must equivalently be one of
the forms\begin{equation} \{ xxxx, yyxx, zzxx, xxyy, xxzz, yyyy,
zzzz, yyzz, \star \star \star \star \}, \tag{5.3}\label{eq:5-3}
\end{equation}\begin{equation} \{ xxxx, yyxx, yxyx, xxyy, yxxy,
xyyx, xyxy, yyyy, \star \star \star \star \},
\tag{5.4}\label{eq:5-4}
\end{equation}\begin{equation} \{ xxxx, yyxx, yxyx, xxyy, yxxy,
xyyx, xyxy, zzzz, \star \star \star \star \}.
\tag{5.5}\label{eq:5-5}
\end{equation}
From Eq.(5.3) we obtain Eq.(5.1), as well from Eqs.(5.4) and (5.5)
obtain Eq.(5.2).

On the other hand, it is evident that one cannot add a element
into Eq.(5.1) or (5.2) for obtaining a larger GHZ-Mermin
experiment. This completes the proof.

{\it Theorem: Nontrivial $\mathrm{GHZ-Mermin}$ experiments of nine
elements for the four-qubit system must equivalently be
$\mathrm{Eq.(5.2)}.$ Moreover, the associated states exhibiting an
``all versus nothing" contradiction between quantum mechanics and
$\mathrm{EPR}$'s local realism are $\mathrm{GHZ}$ states.}

{\it Proof}.~~By the above Proposition, it suffices to show that
Eq.(5.1) is trivial, while Eq.(5.2) is nontrivial.

Indeed, since $(xxxx)\times (yyxx) \times (xxyy) \times (yyyy) =
1, (xxxx)\times (zzxx) \times (xxzz) \times (zzzz) = 1,
(xxxx)\times (yyxx) \times (xxzz) \times (yyzz) = 1,$ and
$(xxxx)\times (zzxx) \times (xxyy) \times (zzyy) = 1,$ for
Eq.(5.1) we have $\varepsilon_1 \varepsilon_2 \varepsilon_4
\varepsilon_6 =1, \varepsilon_1 \varepsilon_3 \varepsilon_5
\varepsilon_7 =1, \varepsilon_1 \varepsilon_2 \varepsilon_5
\varepsilon_8 =1,$ and $\varepsilon_1 \varepsilon_3 \varepsilon_4
\varepsilon_9 =1.$ Then, one can assign $\nu (x_1) =
\varepsilon_1, \nu (y_1) = \varepsilon_2, \nu (z_1) =
\varepsilon_3, \nu (y_3) = \varepsilon_1 \varepsilon_4, \nu (z_3)
= \varepsilon_1 \varepsilon_5,$ and the remaining ones $v( \cdot )
=1.$

On the other hand, Eq.(4D.2) is included in Eq.(5.2) and $|
\varphi \rangle = \frac{1}{\sqrt{2}} \left ( |0000 \rangle - |1111
\rangle \right )$ is a common eigenstate of Eq.(5.2), it is
concluded that Eq.(5.2) is nontrivial. Moreover, as shown in
Section IV.D that the GHZ state is the unique state with
equivalence up to a local unitary transformation which presents
the GHZ-Mermin proof in Eq.(4D.2), we conclude the same result for
Eq.(5.2). This completes the proof.

\section{Conclusion}

By using some subtle mathematical arguments, we present a complete
construction of the GHZ theorem for the four-qubit system. Two
geometric invariants play a crucial role in our argument. We have
shown that a GHZ-Mermin experiment of the four-qubit system
contains at most nine elements and a four-qubit GHZ-Mermin
experiment presenting the GHZ-Mermin-like proof contains at least
five elements. We have exhibited all four-qubit GHZ-Mermin
experiments of 3-9 elements.

In particular, we have proved that the four-qubit states
exhibiting 100\% violation between quantum mechanics and EPR's
local realism are equivalent to $| \mathrm{GHZ} \rangle =
\frac{1}{\sqrt{2}} \left ( |0000 \rangle - |1111 \rangle \right )$
up to a local unitary transformation, which maximally violate the
following Bell inequality\begin{equation}\langle \mathcal{B}
\rangle \leq 9, \tag{6.1}\label{eq:6-1}
\end{equation}where $\mathcal{B} = - x_1 x_2 x_3 x_4 + y_1 y_2 x_3 x_4
+ y_1 x_2 y_3 x_4 + x_1 x_2 y_3 y_4 + y_1 x_2 x_3 y_4 + x_1 y_2
y_3 x_4 + x_1 y_2 x_3 y_4 - y_1 y_2 y_3 y_4 + z_1 z_2 z_3 z_4.$ On
the other hand, as shown in Theorem in Sec.V, the state maximally
violating Eq.(6.1) is unique and equal to $| \mathrm{GHZ}
\rangle.$ This yields that the maximal violation of statistical
predictions is equivalent to (and so implies) the violation of
definite predictions between quantum mechanics and EPR's local
realism for the four-qubit system, as similar to the three-qubit
system \cite{Chen}. Therefore, from the view of EPR's local
realism one concludes that the maximally entangled states of four
qubits should be just the GHZ state \cite{GB}. We would like to
expect the same result holds true for $n$ qubits, that is, all
states exhibiting 100\% violation between quantum mechanics and
EPR's local realism must be GHZ's states up to a local unitary
transformation. This will provides a natural definition of GHZ
states and hence clarifies the maximally entangled states of $n$
qubits.

Note that Eq.(6.1) is not a standard Bell inequality which have
two observables at each site \cite{WW-ZB}. From the viewpoint of
GHZ's theorem we need to study Bell inequalities of $n$ qubits in
which measurements on each particle can be chosen among three spin
observables. This should be helpful to reveal the close
relationship among entanglement, Bell inequalities, and EPR's
local realism \cite{ZBLW}. Since the Bell inequalities and GHZ's
theorem are two main theme on the violation of EPR's local
realism, it turns out that GHZ's theorem and Bell-type
inequalities can be used to reveal what the term maximally
entangled states should actually mean in the multipartite and/or
higher dimensional quantum systems.


\end{document}